\documentclass[prd,aps,nofootinbib,superscriptaddress,floatfix]{revtex4}
\pdfoutput=1

\usepackage{epsfig}
\usepackage{amsmath}
\usepackage{dsfont}

\input{epsf}

\begin{document}


\title{Lattice QCD study of the Boer-Mulders effect in a pion}

\author{M.~Engelhardt}
\affiliation{\normalsize\it Department of Physics, New Mexico State University,
Las Cruces, NM 88003, USA}

\author{P.~H\"agler}
\affiliation{\normalsize\it Institut f\"{u}r Theoretische Physik,
Universit\"{a}t Regensburg, Regensburg, Germany}

\author{B.~Musch}
\affiliation{\normalsize\it Institut f\"{u}r Theoretische Physik,
Universit\"{a}t Regensburg, Regensburg, Germany}

\author{J.~Negele}
\affiliation{\normalsize\it Center for Theoretical Physics, Massachusetts
Institute of Technology, Cambridge, MA 02139, USA}

\author{A.~Sch\"afer}
\affiliation{\normalsize\it Institut f\"{u}r Theoretische Physik,
Universit\"{a}t Regensburg, Regensburg, Germany}

\begin{abstract}
The three-dimensional momenta of quarks inside a hadron are encoded
in transverse momentum-dependent parton distribution functions (TMDs).
This work presents an exploratory lattice QCD study of a TMD observable
in the pion describing the Boer-Mulders effect, which is related to
polarized quark transverse momentum in an unpolarized hadron. Particular
emphasis is placed on the behavior as a function of a Collins-Soper
evolution parameter quantifying the relative rapidity of the struck
quark and the initial hadron, e.g., in a semi-inclusive deep inelastic
scattering (SIDIS) process. The lattice calculation, performed at
the pion mass $m_{\pi } = 518\, \mbox{MeV} $, utilizes a definition
of TMDs via hadronic matrix elements of a quark bilocal operator
with a staple-shaped gauge connection; in this context, the evolution
parameter is related to the staple direction. By parametrizing the
aforementioned matrix elements in terms of invariant amplitudes,
the problem can be cast in a Lorentz frame suited for the lattice
calculation. In contrast to an earlier nucleon study, due to the
lower mass of the pion, the calculated data enable quantitative
statements about the physically interesting limit of large relative
rapidity. In passing, the similarity between the Boer-Mulders effects
extracted in the pion and the nucleon is noted.
\end{abstract}

\maketitle

\section{Introduction}
Transverse momentum-dependent parton distributions (TMDs) \cite{Boer:2011fh}
constitute one of the pillars on which the three-dimensional tomography of
hadrons rests. Together with the three-dimensional spatial information
derived from generalized parton distributions (GPDs), they permit a
comprehensive reconstruction of hadron substructure and thus
have a bearing on seminal topics in hadron physics, such as
orbital angular momentum contributions to nucleon spin, or
spin-orbit correlations in hadrons. Through the selection of
particular parton spin and transverse momentum components, a
variety of effects can be probed, including naively time-reversal odd
(T-odd) quantities such as the Sivers and Boer-Mulders functions;
these only exist by virtue of initial or final state interactions
in corresponding physical processes, introducing a preferred chronology
in the description of the process. For example, in semi-inclusive
deep inelastic scattering (SIDIS), the operative element is final-state
interactions between the struck quark and the hadron remnant; on
the other hand, in the Drell-Yan (DY) process, initial state interactions
before the lepton pair production enable T-odd effects. TMDs thus
in general have to be considered in the context of specific physical
processes, within a factorization framework appropriate for the
process in question, separating the hard reaction from the TMD and
other elements such as fragmentation functions. In the case of T-odd
effects, the process-dependence manifests itself in the prediction
of a sign change of the Sivers and Boer-Mulders functions between
the SIDIS and DY processes \cite{signchange}.

In view of the fundamental importance of TMDs and the rich spectrum of
effects that can be probed, TMDs have been, and continue to be the
target of a variety of experimental efforts. Deep-inelastic scattering
experiments performed by COMPASS \cite{COMPASS}, HERMES \cite{HERMES} and
Jefferson Lab \cite{jlab} have yielded TMD data including evidence
for the T-odd Sivers effect. Complementary Drell-Yan experiments at
COMPASS \cite{compassdy} and Fermilab \cite{fnaldy} are envisaged,
which could, in particular, test the aforementioned sign change between
the SIDIS and DY processes. Related transverse single-spin
asymmetries have been measured at RHIC in polarized proton-proton
collisions \cite{BNL}. Further experimental efforts at RHIC are projected
to provide insight into strong QCD evolution effects expected for the
Sivers TMD \cite{rhic}. TMDs furthermore constitute a central focus
of the proposed Electron-Ion Collider facility \cite{eicwhite}.

To complement these efforts, providing nonperturbative QCD input from
first principles to the analysis of TMD effects, a project to calculate
TMD observables within lattice QCD was initiated and developed in
\cite{straightlett,straightlinks,tmdlat}. The present work constitutes
a continuation of this project. As described in detail below, the formal
definition of TMDs is based on nonlocal operators, specifically quark
bilocal operators with a gauge connection that takes the shape of a staple.
The path followed by the gauge connection is in principle infinite in
length, and thus it cannot be straightforwardly treated in terms of an
operator product expansion, as is commonly done, e.g., for ordinary parton
distributions or generalized parton distributions. In view of this genuinely
nonlocal character of the operators, lattice QCD explorations of
corresponding hadronic matrix elements directly at the nonlocal operator
level were undertaken in \cite{straightlett,straightlinks}, concentrating
initially on the simpler case of straight gauge links connecting the quark
operators. The nonlocal nature of the operators in particular raises novel
questions regarding regularization and renormalization, which were
addressed in considerable detail in \cite{straightlinks}. Whereas
these questions deserve further study, the aforementioned explorations
suggest that it is a viable working assumption to treat nonlocal lattice
operators in analogy to the fashion in which they are treated in continuum
QCD \cite{collbook}, namely, by absorbing divergences into multiplicative
soft factors. These soft factors can then be canceled in appropriate
ratios; this scheme was used to construct TMD observables in the
subsequent investigation \cite{tmdlat}, and will be used in the
present work. Formally related studies of nonlocal lattice operators,
in which a gauge link in the (Euclidean) time direction originates from
the propagation of a heavy auxiliary quark, have been carried out in
\cite{detmold}; also, a direct approach to light-cone distribution
amplitudes based on nonlocal lattice operators was laid out in
\cite{vbraun}. Moreover, the comprehensive framework for investigating
parton physics within lattice QCD put forward in \cite{Ji_1} and
developed and explored in \cite{Ji_2,Ji_3,steffens} relies
on a direct treatment of such nonlocal lattice operators.

The present work focuses on a TMD observable related to the Boer-Mulders
effect in a pion. Lattice QCD studies of pion structure, predominantly
focusing on form factors, have been previously reported in
\cite{bonnet,ff1,ff2,pion_GPD,ff3,ff4,ff5,ff6,ff7}. Choosing the pion as
the hadron state is motivated by the principal goal of the investigation
presented here, namely, understanding the behavior of TMDs as a function
of an evolution parameter quantifying the rapidity difference between
the hadron momentum and a vector describing the trajectory of the struck
quark. Details are furnished further below. In the previous nucleon
study \cite{tmdlat}, no definite conclusions regarding the limit of
large rapidity difference proved possible. By virtue of its lower mass,
the pion provides a larger rapidity difference at given momentum, and
this choice of hadron state thus aids in approaching the limit of
physical interest. In addition, the spinless nature of the pion permits
additional spatial averaging to suppress statistical uncertainties.
Indeed, the chief advance of the present work lies in providing
quantitative insight into the limit of large evolution parameter.
Preliminary accounts of this work were given in
\cite{latt13,qcdevol14,lc2014}.

\section{Definition of TMD observables}
\subsection{Correlation functions}
\begin{figure}[b]
\centerline{\psfig{file=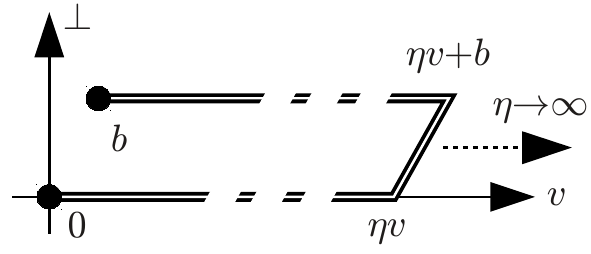,width=7cm }}
\caption{Gauge link structure ${\cal U} [0,\eta v, \eta v+b,b]$ in the
correlator (\ref{spacecorr}). In a concrete lattice calculation, the
limit $\eta \rightarrow \infty $ has to be taken numerically.}
\label{fig_2a}
\end{figure}

Quark transverse momentum-dependent parton distributions (TMDs) can be
defined in terms of the fundamental correlator
\begin{equation}
\widetilde{\Phi }^{[\Gamma ]}_{\mbox{\scriptsize unsubtr.} } (b,P,\ldots )
\equiv \frac{1}{2} \langle P | \ \bar{q} (0) \
\Gamma \ {\cal U} [0,\eta v, \eta v+b,b] \ q(b) \ |P\rangle \ \ ,
\label{spacecorr}
\end{equation}
where $P$ denotes the momentum of the hadron state; the present work
focuses on pions, and thus no spin is attached to the state. $\Gamma $
represents an arbitrary Dirac $\gamma $ matrix structure. The quark
operators at positions $0$ and $b$ are connected by the gauge link
${\cal U} [0,\eta v, \eta v+b,b]$, which connects the points listed in
its argument by straight-line segments; thus, the gauge link has the
shape of a staple, cf.~Fig.~\ref{fig_2a}, with the unit vector $v$
specifying the staple direction and $\eta $ its length. One is
ultimately interested in the limit $\eta \rightarrow \infty $, which
in a concrete lattice calculation is of course reached by extrapolation.
This gauge link form incorporates final state interactions between the
struck quark and the hadron remnant in semi-inclusive deep inelastic
scattering (SIDIS) \cite{pijlman}, and analogously initial state
interactions in the Drell-Yan process (DY).
The ellipsis in the argument of
$\widetilde{\Phi }^{[\Gamma ]}_{\mbox{\scriptsize unsubtr.} } $ indicates
that the correlator will depend on various further parameters, related,
e.g., to regularization, specified below as needed.

Fourier transformation of (\ref{spacecorr}),
\begin{eqnarray}
\Phi^{[\Gamma ]} (x,k_T ,P,\ldots ) &=&
\int \frac{d^2 b_T }{(2\pi )^2 } \int \frac{d(b\cdot P)}{(2\pi )P^{+} }
\exp \left( ix(b\cdot P) -ib_T \cdot k_T \right)
\left. \frac{\widetilde{\Phi }^{[\Gamma ]}_{\mbox{\scriptsize unsubtr.} }
(b,P,\ldots )}{\widetilde{\cal S} (b_T^2 ,\ldots )}
\ \right|_{b^{+} =0} \ \ \ \ \ \ \ \ \,
\label{fundcorr}
\end{eqnarray}
leads to the momentum space correlator $\Phi^{[\Gamma ]} $, which
ultimately will be parametrized in terms of TMDs, cf.~below. The
position space correlator (\ref{spacecorr}), as written, requires
regularization not only of the quark operator self-energies, but also
of the self-energy of the Wilson line (this is indicated by the
subscript ``unsubtr.''). The regularization of the latter is effected
by dividing by the soft factor $\widetilde{\cal S} $. The detailed
structure of the soft factor depends on the concrete factorization
approach employed. For example, in the scheme developed in
\cite{collbook,spl1}, which, for reasons discussed further below,
provides the phenomenological framework for the present study, the
soft factor takes the form
\begin{equation}
\widetilde{\cal S} (b_T^2 ,\ldots ) =
\sqrt{\frac{\tilde{S}_{(0)} (b_T , +\infty , -\infty )
\tilde{S}_{(0)} (b_T , y_s , -\infty )}{\tilde{S}_{(0)}
(b_T , +\infty , y_s )} }
\end{equation}
where $\tilde{S}_{(0)} (b_T ,\ldots )$ is a vacuum expectation
value of Wilson line structures extending, initially, into space-like
directions; some of them remain at finite rapidity $y_s $, whereas
others are taken to the light-cone limit, i.e., infinite rapidity.
This particular form of the soft factor, containing more than one
rapidity, cannot be cast in a Lorentz frame in which it exists at
a single time, and for this reason, it is not suited for evaluation
within lattice QCD. However, the observables that will be defined
further below are ratios in which the soft factors cancel \cite{tmdlat}.
There is, therefore, no obstacle to evaluating those observables
within lattice QCD, and the detailed form of the soft factor is
immaterial. An alternative construction of TMD soft factors which,
in principle, is amenable to lattice QCD evaluation has been put
forward in \cite{Ji_3} within the framework laid out in
\cite{Ji_1}.

As written in (\ref{fundcorr}), the transverse components $b_T $
of the quark separation $b$ are Fourier conjugate to the quark transverse
momentum $k_T $, while the longitudinal component $b\cdot P$ is
Fourier conjugate to the longitudinal momentum fraction $x$. The present
work will be confined to the case $b\cdot P=0$, corresponding to evaluating
the integral with respect to $x$ over the correlator $\Phi^{[\Gamma ]} $
and TMDs derived from it. It should be stressed, however, that there is
no obstacle to extending calculations of the type presented here to a
scan of the $b\cdot P$-dependence\footnote{In a practical calculation,
the range of accessible $b\cdot P$ is limited by the available $b$ and
$P$, $|b\cdot P| \leq |{\bf P} | \sqrt{-b^2 } $ (where ${\bf P} $ denotes
the spatial momentum), leading to an
increasing systematic uncertainty at small $x$.}, yielding upon
Fourier transformation the $x$-dependence of $\Phi^{[\Gamma ]} $ and
the TMDs under consideration. Studies of the $b\cdot P$-dependence in
the straight gauge link ($\eta v=0$) case have already been carried
out in \cite{straightlinks}, and further investigations in this direction
are planned for future work. A related proposal to obtain the $x$-dependence
of parton distributions has been put forward and explored in
\cite{Ji_1,Ji_2,Ji_3,steffens}.

Finally, (\ref{fundcorr}) is evaluated at $b^{+} =0$, in accordance
with the standard phenomenological framework, which employs a Lorentz frame
in which the hadron of mass $m_h $ propagates with a large momentum in
3-direction, $P^{+} \equiv (P^0 +P^3 )/\sqrt{2} \gg m_h $; then, the
quark momentum components scale such that the correlator (\ref{fundcorr})
and TMDs derived from it are principally functions of the quark
longitudinal momentum fraction $x=k^{+} /P^{+} $ and the quark
transverse momentum vector $k_T $, whereas the dependence on the
component $k^{-} \equiv (k^0 -k^3 )/\sqrt{2} \ll m_h $ becomes irrelevant
in this limit. Correspondingly, (\ref{fundcorr}) is regarded as having been
integrated over $k^{-} $, implying that the conjugate variable $b^{+} $
is to be set to zero, as written.

Before continuing with the parametrization of $\Phi^{[\Gamma ]} $ in
terms of TMDs, it is important to note a further specification with
respect to the staple direction $v$. The Wilson lines along the legs of
the staple represent an effective, resummed description of gluon exchanges
between, in the case of SIDIS, the struck quark and the hadron remnant
\cite{pijlman}. Accordingly, their direction $v$ should be taken to follow
the path of the ejected quark, close to the light cone from the point of
view of the hadron. Whereas, at tree level, there is no obstacle to the
most straightforward choice, namely, a light-like $v$, beyond tree level,
this choice is associated with rapidity divergences \cite{rapidrev}.
Various schemes have been advanced to treat these divergences
\cite{collbook,spl1,Ji:2004wu,idilbi}, and the equivalence between
some of them discussed in \cite{tmdequiv}. In particular, the
scheme advanced in \cite{collbook,spl1} effects regularization by
tilting the staple direction $v$ slightly off the light cone into
the spacelike region. This feature is crucial for a concrete
implementation of a lattice QCD evaluation of matrix elements of
the type (\ref{spacecorr}), as will be detailed further below.
Thus, the phenomenological framework providing the backdrop for the
present treatment is, specifically, the one advanced in \cite{collbook,spl1}.
Choosing $v$ to be spacelike for the purposes of regularization implies
a dependence of the calculation on an additional parameter, characterizing
proximity of the staple to the light cone. Here, this parameter will be
chosen as
\begin{equation}
\hat{\zeta} = \frac{v\cdot P}{|v| |P| } \ ,
\label{zetahat}
\end{equation}
where the absolute length of a four-vector $w$ is denoted by
$|w|=\sqrt{| w^2 |} $. In terms of $\hat{\zeta} $, the light-cone limit
corresponds to $\hat{\zeta} \rightarrow \infty $. The generated lattice
data thus have to be extrapolated to the double limit
$\eta \rightarrow \infty $, $\hat{\zeta} \rightarrow \infty $,
i.e., the limit of infinite staple length, and staple direction
converging toward the light cone.

Alternative to this purely kinematic characterization, also the parameter
$\zeta = 2m_h \hat{\zeta} $ is frequently employed \cite{cs},
and viewed as a dynamical scale, to be compared to $\Lambda_{QCD} $;
perturbative evolution equations in $\zeta $ can then be derived at
sufficiently large $\zeta $, cf., e.g., \cite{spl2}. The question
whether one has reached the asymptotic regime appropriate for the
definition of TMDs presumably has both kinematic and dynamical aspects.
The applicability of perturbation theory, e.g., for determining
evolution equations is a dynamical issue most adequately characterized
by considering the dimensionful parameter $\zeta $ in relation to the
characteristic QCD scale, i.e., requiring $\zeta \gg \Lambda_{QCD} $.
On the other hand, this by itself does not necessarily guarantee
kinematics close to the light cone, $\hat{\zeta} \gg 1$. Presumably,
both conditions need to be taken into account in general. In the
present work, no quantitative connection to perturbative evolution is
attempted; the numerical data do not reach values of $\zeta $
which lie clearly within the perturbative regime. The maximum spatial
momentum employed is $1.17\, \mbox{GeV} $, and the maximum value of
$\zeta $ is $\zeta =2.1\, \mbox{GeV} $, corresponding to
$\hat{\zeta} =2.03$. Instead, the dependence of the lattice data on the
kinematical variable $\hat{\zeta} $ will be studied empirically, including
exploration of ad hoc ans\"atze for the large-$\hat{\zeta} $
behavior, allowing for corresponding extrapolations. Characterizing
the large-$\hat{\zeta} $ limit is in fact the primary goal of the present
investigation. This limit was seen to present a considerable challenge
in the previous study \cite{tmdlat}, which considered nucleon TMDs;
no definite statements concerning the large-$\hat{\zeta} $ behavior
proved possible. The present work, focusing on pions, permits accessing
higher $\hat{\zeta} $ both by virtue of the lighter hadron mass
(note that the hadron mass enters the denominator of (\ref{zetahat})),
and by employing additional spatial averaging facilitated by the spinless
nature of the pion, enhancing statistics. As will be seen further below,
the data extracted in the present investigation are of sufficient
quality to yield a signal for the $\hat{\zeta} \rightarrow \infty $
limit of the generalized Boer-Mulders shift defined in eq.~(\ref{gbmshift})
below. This constitutes the main advance of this work.

\subsection{Parametrizations}
Returning to the momentum space correlator $\Phi^{[\Gamma ]} $, its
parametrization in terms of the relevant Lorentz structures yields,
at leading twist,
\begin{eqnarray}
\Phi^{[\gamma^{+} ]} (x,k_T ,P,\ldots ) &=&
f_1 (x,k_T ,P,\ldots ) \label{tmdunpol} \\
\Phi^{[i\sigma^{i+} \gamma^{5} ]} (x,k_T ,P,\ldots ) &=&
-\frac{\epsilon^{-+ij} k_j }{m_{\pi } }
h_1^{\perp } (x,k_T ,P,\ldots ) \ . \label{tmdbm}
\end{eqnarray}
For spinless particles such as the pion, there are only two leading twist
TMDs, in contrast to the eight which arise for spin-$\frac{1}{2} $
particles \cite{tmd1,tmd2,tmd3}. The TMD $f_1 $ is simply the unpolarized
quark distribution, whereas the Boer-Mulders function \cite{bm}
$h_1^{\perp } $ encodes the distribution of transversely polarized
quarks in the pion. The Boer-Mulders function is odd under time
reversal (T-odd). Physically, it only exists by virtue of the final
and initial state interactions, in the SIDIS and DY processes,
respectively, which break the symmetry of the processes under time
reversal. Formally, it is the introduction of the additional vector
$v$ describing the staple direction in the staple-shaped gauge link
which breaks the symmetry; the Boer-Mulders function vanishes for
a straight gauge link, $\eta v=0$. Correspondingly, the generalized
Boer-Mulders shift defined in eq.~(\ref{gbmshift}) below will be
an odd function of $v$.

On the other hand, one can also decompose the position space correlator
$\widetilde{\Phi }^{[\Gamma ]}_{\mbox{\scriptsize unsubtr.} } $ into
invariant amplitudes \cite{tmdlat}. The general decomposition
is (the combinations corresponding specifically to the leading twist
TMDs (\ref{tmdunpol}) and (\ref{tmdbm}) will be considered further
below):
\begin{eqnarray}
\frac{1}{2} \widetilde{\Phi }^{[\mathds{1}]}_{\mbox{\scriptsize unsubtr.} }
&=& m_{\pi } \widetilde{A}_{1} \\
\frac{1}{2} \widetilde{\Phi }^{[\gamma^{\mu }]}_{\mbox{\scriptsize unsubtr.} }
&=& P^{\mu } \widetilde{A}_{2} -im_{\pi }^{2} b^{\mu } \widetilde{A}_{3}
+\frac{m_{\pi }^{2} }{v\cdot P} v^{\mu } \widetilde{B}_{1} \\
\frac{1}{2}
\widetilde{\Phi }^{[\gamma^{\mu } \gamma^{5} ]}_{\mbox{\scriptsize unsubtr.} }
&=& \frac{im_{\pi }^{2} }{v\cdot P} \epsilon^{\mu \nu \rho \sigma } P_{\nu }
b_{\rho } v_{\sigma } \widetilde{B}_{4} \\
\frac{1}{2}
\widetilde{\Phi }^{[i\sigma^{\mu \nu }
\gamma^{5} ]}_{\mbox{\scriptsize unsubtr.} }
&=& im_{\pi } \epsilon^{\mu \nu \rho \sigma } P_{\rho } b_{\sigma }
\widetilde{A}_{4} - \frac{m_{\pi } }{v\cdot P} \epsilon^{\mu \nu \rho \sigma }
P_{\rho } v_{\sigma } \widetilde{B}_{2} + \frac{im_{\pi }^{3} }{v\cdot P}
\epsilon^{\mu \nu \rho \sigma } b_{\rho } v_{\sigma } \widetilde{B}_{3}
\end{eqnarray}
The present treatment focuses on the special case $b\cdot P=0$, which
in the context of TMDs, defined in a frame in which $b^{+} =0$ and
$v_T = P_T =0$, also implies $b\cdot v=0$ \cite{tmdlat}.
Under these constraints, the above relations are readily inverted. The
amplitudes needed below are, explicitly,
\begin{eqnarray}
\widetilde{A}_{2} &=& \frac{1}{1+\hat{\zeta}^{2} } \frac{1}{2m_{\pi }^{2} }
\left( P_{\mu } - \frac{v\cdot P}{v^2 } v_{\mu } \right)
\widetilde{\Phi }^{[\gamma^{\mu }]}_{\mbox{\scriptsize unsubtr.} }
\label{ampa2} \\
\widetilde{B}_{1} &=& \frac{\hat{\zeta}^{2} }{1+\hat{\zeta}^{2} }
\frac{1}{2m_{\pi }^{2} }
\left( P_{\mu } - \frac{m_{\pi }^{2} }{v\cdot P} v_{\mu } \right)
\widetilde{\Phi }^{[\gamma^{\mu }]}_{\mbox{\scriptsize unsubtr.} }
\label{ampb1} \\
\widetilde{A}_{4} &=& i\frac{1}{1+\hat{\zeta}^{2} }
\frac{1}{4b^2 m_{\pi }^{3} }
\left( P^{\kappa } - \frac{v\cdot P}{v^2 } v^{\kappa } \right)
b^{\lambda } \epsilon_{\kappa \lambda \mu \nu }
\widetilde{\Phi }^{[i\sigma^{\mu \nu }
\gamma^{5} ]}_{\mbox{\scriptsize unsubtr.} } \\
\widetilde{B}_{3} &=& -i\frac{\hat{\zeta}^{2} }{1+\hat{\zeta}^{2} }
\frac{1}{4b^2 m_{\pi }^{3} }
\left( P^{\kappa } - \frac{m_{\pi }^{2} }{v\cdot P} v^{\kappa } \right)
b^{\lambda } \epsilon_{\kappa \lambda \mu \nu }
\widetilde{\Phi }^{[i\sigma^{\mu \nu }
\gamma^{5} ]}_{\mbox{\scriptsize unsubtr.} }
\label{ampb3}
\end{eqnarray}
Note that $\widetilde{B}_{1} $ and $\widetilde{B}_{3} $ are regular for
$v\cdot P \rightarrow 0$ owing to the $\hat{\zeta}^{2} $ prefactor.
Of particular interest are the leading twist objects
\begin{eqnarray}
\frac{1}{2P^{+} }
\widetilde{\Phi }^{[\gamma^{+} ]}_{\mbox{\scriptsize unsubtr.} } &=&
\widetilde{A}_{2B} \label{amp2b} \\
\frac{1}{2P^{+} }
\widetilde{\Phi }^{[i\sigma^{i+} \gamma^{5} ]}_{\mbox{\scriptsize unsubtr.} }
&=& -im_{\pi } \epsilon^{-+ij} b_j \widetilde{A}_{4B}
\label{amp4b}
\end{eqnarray}
(where $i,j$ denote transverse spatial indices),
given in terms of the amplitude combinations
\begin{eqnarray}
\widetilde{A}_{2B} &=& \widetilde{A}_{2} +
\left( 1-\sqrt{1+\hat{\zeta }^{-2} } \right) \widetilde{B}_{1} \\
\widetilde{A}_{4B} &=& \widetilde{A}_{4} -
\left( 1-\sqrt{1+\hat{\zeta }^{-2} } \right) \widetilde{B}_{3}
\label{a4bdef}
\end{eqnarray}
Also these combinations are regular for $\hat{\zeta} \rightarrow 0$ by
virtue of the $\hat{\zeta}^{2} $ prefactors in (\ref{ampb1}),(\ref{ampb3}).
Note that, in the case of vanishing spatial momentum, one cannot identify
``forward'' and ``backward'' directions for $v$; there is then only a
single branch in $|\eta v|$, the sign of which is a matter of definition.
Although only the combinations $\widetilde{A}_{2B} $ and
$\widetilde{A}_{4B} $ appear in (\ref{amp2b}),(\ref{amp4b}),
for the numerical analysis to follow, it will be valuable
to be able to consider $\widetilde{A}_{2} $, $\widetilde{B}_{1} $,
$\widetilde{A}_{4} $ and $\widetilde{B}_{3} $ individually, not just
those combinations.

Given that $\widetilde{\Phi }^{[\Gamma ]}_{\mbox{\scriptsize unsubtr.} } $
and $\Phi^{[\Gamma ]} $ are related via a Fourier transformation,
the quantities arising in the respective decompositions
(\ref{tmdunpol})-(\ref{tmdbm}) and (\ref{amp2b})-(\ref{amp4b}) must
be similarly related, i.e., the amplitudes $\widetilde{A}_{iB} $
must be related to Fourier-transformed TMDs. Indeed, denoting
$x$-moments of generic Fourier-transformed TMDs by
\begin{eqnarray}
\tilde{f}^{[m](n)} (b_T^2 ,\ldots ) &=& n! \left( -\frac{2}{m_h^2 }
\partial_{b_T^2 } \right)^{n} \int_{-1}^{1} dx\, x^{m-1}
\int d^2 k_T \,
e^{ib_T \cdot k_T } \ f(x,k_T^2 ,\ldots )
\\
&=& 
\frac{2\pi n!}{(m_h^2)^n}\, \int_{-1}^{1} dx\, x^{m-1}
\int d |k_T | \, |k_T | \,
\left( \frac{|k_T |}{|b_T |} \right)^n 
J_n(|b_T | |k_T |) f(x, k_T^2, \ldots )
\label{eq10}
\end{eqnarray}
where $J_n $ denotes the Bessel functions of the first kind, one finds
\cite{tmdlat}
\begin{eqnarray}
\tilde{f}_{1}^{[1](0)} (b_T^2 , \hat{\zeta } ,\ldots , \eta v\cdot P)
&=& 2\widetilde{A}_{2B} (-b_T^2 ,b\cdot P=0, b\cdot v=0,
\hat{\zeta } ,\eta v\cdot P) / \widetilde{\cal S} (b_T^2 ,\ldots )
\label{ampdec1} \\
\tilde{h}_{1}^{\perp [1](1)} (b_T^2 ,\hat{\zeta } ,\ldots ,\eta v\cdot P)
&=& 2\widetilde{A}_{4B} (-b_T^2 ,b\cdot P=0, b\cdot v=0,
\hat{\zeta } ,\eta v\cdot P) / \widetilde{\cal S} (b_T^2 ,\ldots )
\label{ampdec2}
\end{eqnarray}
Note the appearance of the soft factors on the right hand sides.

\subsection{Boer-Mulders shift}
As already indicated further above, one obtains an observable in which
the soft factors cancel by forming a suitable ratio, namely, the
``generalized Boer-Mulders shift''
\begin{equation}
\langle k_y \rangle_{UT} (b_T^2 , \ldots ) \equiv m_{\pi }
\frac{\tilde{h}_{1}^{\perp [1](1)}
(b_T^2 ,\ldots )}{\tilde{f}_{1}^{[1](0)} (b_T^2 , \ldots )}
= m_{\pi } \frac{\widetilde{A}_{4B} (-b_T^2 ,0,0,\hat{\zeta } ,
\eta v\cdot P)}{\widetilde{A}_{2B} (-b_T^2 ,0,0,\hat{\zeta } ,
\eta v\cdot P)}
\label{gbmshift}
\end{equation}
Note that ratios of this type also cancel $\Gamma $-independent
multiplicative field renormalization constants attached to the quark
operators in (\ref{spacecorr}) at finite physical separation $b$.
It should be emphasized that the construction thus far is a
continuum perturbative QCD construction. That this construction
carries across into lattice QCD, i.e., that the lattice operators
are similarly regularized and renormalized by multiplicative soft
factors which cancel in ratios, is a working assumption which was
already explored in considerable detail in \cite{straightlinks},
and which will be investigated further in future work. Physically,
this assumption appears plausible at least at separations substantially
larger than the lattice spacing, where the lattice operators are
expected to approximate the corresponding continuum operators.

To interpret the generalized Boer-Mulders shift, note that
the $b_T \rightarrow 0$ limit of the quantities defined in
(\ref{eq10}) formally corresponds to $k_T^{2} $-moments of TMDs,
\begin{eqnarray}
\tilde{f}^{[m](n)} (0,\ldots ) &=& \int_{-1}^{1} dx\, x^{m-1}
\int d^2 k_T \ \left( \frac{k_T^2 }{2m_h^2 }\right )^{n} \
f(x,k_T^2 ,\ldots )
\label{ktmom}
\end{eqnarray}
Thus, in the formal $b_T \rightarrow 0$ limit, the generalized
Boer-Mulders shift reduces to the ``Boer-Mulders shift''
\begin{equation}
\langle k_y \rangle_{UT} (0, \ldots ) = 
m_{\pi } \frac{\tilde{h}_{1}^{\perp [1](1)}
(0, \ldots )}{\tilde{f}_{1}^{[1](0)} (0, \ldots )} =
\left. \frac{\int dx \int d^2 k_T \, k_y
\Phi^{[\gamma^{+} + s^j i\sigma^{j+} \gamma^{5} ]}
(x,k_T ,P,\ldots )}{\int dx \int d^2 k_T \,
\Phi^{[\gamma^{+} + s^j i\sigma^{j+} \gamma^{5} ]}
(x,k_T ,P,\ldots )} \right|_{s_T =(1,0)}
\label{bmshift}
\end{equation}
which, in view of the structure of the right-hand side, formally takes the
form of the average transverse momentum in $y$-direction of quarks
polarized in the transverse (``$T$'') $x$-direction, in an unpolarized
(``$U$'') pion, normalized to the corresponding number of valence quarks.
This is in accord with the $({\bf s}_{T} \times {\bf b}_{T})\cdot {\bf P} $
structure of the correlator (\ref{amp4b}), where ${\bf s}_{T} $ represents
the quark spin. The numerator in (\ref{bmshift}) sums over the contributions
from quarks and antiquarks, whereas the denominator contains the difference
between quark and antiquark contributions, thus giving the number of valence
quarks \cite{straightlinks,muldtan}. It should be noted, however, that the
$k_T^{2} $-moments of TMDs (\ref{ktmom}) appearing in (\ref{bmshift})
are in general divergent \cite{bacchetta} at large $k_T $ and thus
not well-defined absent further regularization. The generalized quantity
(\ref{gbmshift}) is a natural regularization, with finite $b_T $
effectively acting as a regulator through the associated Bessel weighting,
cf.~(\ref{eq10}). This Bessel weighting also is advantageous in the
analysis of experimental asymmetries \cite{bweight1,bweight2}. In the
present work, lattice QCD data for the generalized Boer-Mulders shift
(\ref{gbmshift}) will be obtained and presented at finite $b_T $.
The path by which these data can be obtained proceeds via lattice QCD
evaluation of the fundamental correlator (\ref{spacecorr}) for a range
of Dirac and staple link structures, extraction of the relevant invariant
amplitudes (\ref{ampa2})-(\ref{ampb3}), and construction of the ratio
(\ref{gbmshift}). As already mentioned further above, for this lattice
QCD calculational scheme to be viable, it is necessary to employ a
phenomenological framework such as the one advanced in \cite{collbook,spl1},
in which all separations in the correlator (\ref{spacecorr}) are spacelike,
including the staple direction $v$. Such a scheme implies dependence on
the Collins-Soper-type evolution parameter $\hat{\zeta }$,
cf.~(\ref{zetahat}), quantifying proximity of the staple to the
light cone. The principal focus of the present investigation is,
indeed, the dependence of the generalized Boer-Mulders shift on
$\hat{\zeta }$, including its asymptotic $\hat{\zeta } \rightarrow \infty $
behavior.

\section{Lattice QCD calculations}
Lattice QCD employs a Euclidean time coordinate, serving to project
out hadronic ground states via the associated exponentially decaying
time evolution. As a consequence, when evaluating matrix elements of
operators in hadronic states, no Minkowski time separations in those
operators can be accomodated; one is restricted to operators which
are defined at one single time. This is the reason why it is imperative
to employ a framework in which all separations in the fundamental
correlator (\ref{spacecorr}) are spacelike. Only in this case is
there no obstacle to boosting the problem to a Lorentz frame in which
the operator in (\ref{spacecorr}) exists at a single time, and
performing the lattice calculation in that particular frame.

The decomposition of the resulting correlators into invariant
amplitudes, cf.~(\ref{ampa2})-(\ref{ampb3}), is a further crucial
element of the present treatment. Expressed in this fashion, the
results of the lattice calculation are immediately applicable also in
the original Lorentz frame in which (\ref{spacecorr}) was initially
defined. Finally, as already discussed above, the construction of
ratios of amplitudes in which soft factors cancel serves to connect
the results to phenomenological observables such as the generalized
Boer-Mulders shift (\ref{gbmshift}).

The lattice QCD data for the present exploration were obtained within a
mixed action scheme employing domain wall valence quarks on an $N_f =2+1$
dynamical asqtad quark gauge ensemble provided by the MILC collaboration
\cite{milc}. Since the principal focus lies on understanding the systematics
of the large $\hat{\zeta } $ limit, which proved inaccessible in previous
investigations, a fairly high pion mass, $m_{\pi}=518\, \mbox{MeV} $, was
chosen for this study to alleviate statistical fluctuations. Further
details of the ensemble are given in Table~\ref{Tab_1}.
\begin{table}
\centering
\begin{tabular}{|c|c|c|c|c|c|c|c|}
\hline
$L^3\times T$ & $a$(fm) & $am_{u,d}$ & $am_s$ &
$m_{\pi}^{\mbox{\tiny DWF} }$ (MeV) & $m_N^{\mbox{\tiny DWF} } $ (GeV) &
\#conf. & \#meas. \\
\hline
$20^3\times 64$ & 0.11849(14)(99) & 0.02  & 0.05 &
518.4(07)(49) & 1.348(09)(13) & 486 & 3888 \\
\hline
\end{tabular}
\caption{Parameters of the lattice ensemble. Note that the lattice spacing
$a$ was determined in a different scheme than in \cite{LHPC_1,LHPC_2}, and
consequently it, as well as the pion and nucleon masses quoted, differ
slightly from the ones given in the aforementioned references \cite{tmdlat}.
The first error quoted for the pion and nucleon masses is statistical, the
second stems from the conversion to physical units using $a$. The bare
asqtad quark masses are denoted $m_{u,d,s} $. Eight measurements were
made on each gauge configuration.}
\label{Tab_1}
\end{table}
This mixed action scheme, including the specific ensemble employed
here, has been used extensively by the LHP Collaboration for studies
of hadron structure, cf., e.g., \cite{LHPC_1,LHPC_2}. It also provided
the basis for the previous nucleon TMD investigation reported in
\cite{tmdlat}.

To extract the correlator (\ref{spacecorr}), one evaluates both
three-point functions $C_{\mbox{\tiny 3pt} } $ and two-point
functions $C_{\mbox{\tiny 2pt} } $ with pion sources and sinks
of definite spatial momentum\footnote{In practice, momentum
conservation eliminates the need for projection at the source,
provided one projects onto zero momentum transfer at the
operator insertion in $C_{\mbox{\tiny 3pt} } $ instead.} ${\bf P} $,
\begin{eqnarray}
C_{\mbox{\tiny 3pt} } [\hat{O} ] (t_i ,t,t_f ,P) &=&
\sum_{{\bf x}_{i} ,{\bf x}_{f} }
e^{-i({\bf x}_{f} -{\bf x}_{i} ) \cdot {\bf P} }
\langle \phi (t_f ,{\bf x}_{f} ) \hat{O} (t)
\phi^{\dagger } (t_i ,{\bf x}_{i} ) \rangle \\
C_{\mbox{\tiny 2pt} } (t_i ,t_f ,P) &=&
\sum_{{\bf x}_{i} ,{\bf x}_{f} }
e^{-i({\bf x}_{f} -{\bf x}_{i} ) \cdot {\bf P} }
\langle \phi (t_f ,{\bf x}_{f} )
\phi^{\dagger } (t_i ,{\bf x}_{i} ) \rangle
\end{eqnarray}
where $t_i $, $t$ and $t_f $ are source time, operator insertion time
and sink time, respectively, and $\phi $ denotes an interpolating field
with the quantum numbers of the pion. Wuppertal-smeared quark fields
were employed to construct these pion sources and sinks, which were
separated by $t_f -t_i = 9a$. Only connected contractions contributing to
$C_{\mbox{\tiny 3pt} } $ were evaluated; disconnected contributions, which
are expected to be small, are omitted in all results presented below.
In the case of the $\pi^{+} $ meson, $u$-quark and $\bar{d} $-quark
distributions coincide; contrary to the nucleon case, there is therefore
no nontrivial $u-d$ quark combination in which disconnected contributions
exactly cancel.

The fundamental correlator (\ref{spacecorr}) is then obtained from
plateaus in $t$ for $t_i \ll t \ll t_f $ in the three-point to two-point
function ratio \cite{bonnet},
\begin{equation}
\widetilde{\Phi }^{[\Gamma ]}_{\mbox{\scriptsize unsubtr.} } =
E(P) \frac{C_{\mbox{\tiny 3pt} } [\hat{O} ]
(t_i ,t,t_f ,P)}{C_{\mbox{\tiny 2pt} } (t_i ,t_f ,P)}
\label{3to2rat}
\end{equation}
where $E(P)$ is the energy of the pion state and $\hat{O} $ is taken to
be the operator in (\ref{spacecorr}). It should be noted that, at the
employed source-sink separation of $t_f -t_i = 9a = 1.07\, \mbox{fm} $,
significant excited state contaminations in the plateaus extracted from
(\ref{3to2rat}) cannot be excluded a priori. This issue was not
investigated in the present exploratory study at the fairly high pion mass 
$m_{\pi } =518\, \mbox{MeV} $. However, in future work at lower pion
masses, where excited state contaminations are exacerbated, it will
present an additional challenge.

The set of combinations of pion momenta and staple-shaped gauge link
paths used is listed in Table~\ref{stapletable}. It should be noted
that, in the case of either ${\bf b} $ or ${\bf v} $ extending into a
direction which does not coincide with a lattice axis, there is more
than one optimal approximation of the corresponding continuum path by
a lattice link path; e.g., if ${\bf b} = 2({\bf e}_{1} + {\bf e}_{2} )$,
where ${\bf e}_{i} $ denotes the lattice link vector in $i$-direction,
both the sequence of links
$({\bf e}_{1} ,{\bf e}_{2} ,{\bf e}_{1} ,{\bf e}_{2} )$ and the sequence
of links $({\bf e}_{2} ,{\bf e}_{1} ,{\bf e}_{2} ,{\bf e}_{1} )$ equally
well approximate the continuum path. In such a case,
$\widetilde{\Phi }^{[\Gamma ]}_{\mbox{\scriptsize unsubtr.} } $ was
always averaged over all equivalent lattice link paths, for both
${\bf b} $ and ${\bf v} $ vectors. This symmetry improvement of the
lattice operators is important to preserve the manifest time-reversal
transformation properties present for the continuum staple-shaped gauge
link path operators.

Note also that, in the mixed action scheme used for these calculations,
before evaluating domain wall propagators for valence quarks, the asqtad
gauge configurations are HYP-smeared to reduce dislocations (or rough
fields) that would otherwise allow right-handed states on one domain
wall to mix with left-handed states on the other domain wall. The
lattice gauge link paths in (\ref{spacecorr}) were constructed using
those same HYP-smeared gauge configurations. This has the advantageous
consequence that renormalization constants are closer to their tree-level
values, while it would have no effect in the continuum limit. On the other
hand, as shown in ref.~\cite{steffens}, there are significant differences
between 0 and 2 steps of HYP-smearing in the direct calculation of parton
distributions using straight gauge link paths instead of staples
(see Figs.~3-5 therein), so the optimal use of HYP smearing requires
further study.

\begin{table}
\begin{tabular}{|c|c|c|}
\hline
${\bf b} /a $ & $\eta {\bf v} /a $
& ${\bf P} \cdot aL /(2\pi )$ \\
\hline\hline
$n\cdot (0,0,1), n=-9,\ldots,9$ &
$\pm n^{\prime } \cdot (1,0,0)$
& $(0,0,0)$, $(-1,0,0)$, $(-2,0,0)$, \\
& & $(0,-2,0)$, $(-1,-1,0)$, $(-2,-1,0)$ \\
\cline{2-3}
& $\pm n^{\prime } \cdot (0,1,0)$
& $(0,0,0)$, $(-1,0,0)$, $(-2,0,0)$, \\
& & $(0,-2,0)$, $(-1,-1,0)$, $(-2,-1,0)$ \\
\cline{2-3}
& $\pm n^{\prime } \cdot (1,1,0)$
& $(0,0,0)$, $(-1,0,0)$, $(-2,0,0)$, \\
& & $(0,-2,0)$, $(-1,-1,0)$, $(-2,-1,0)$ \\
\cline{2-3}
& $\pm n^{\prime } \cdot (1,-1,0)$
& $(0,0,0)$, $(-1,0,0)$, $(-2,0,0)$, $(0,-2,0)$, $(-2,-1,0)$ \\
\cline{2-3}
& $\pm n^{\prime } \cdot (2,1,0)$ & $(-2,-1,0)$ \\
\cline{2-3}
& $\pm n^{\prime } \cdot (2,-1,0)$ & $(-2,-1,0)$ \\
\cline{1-3}
$n\cdot (0,1,0), n=-9,\ldots,9$ &
$\pm n^{\prime } \cdot (1,0,0)$
& $(0,0,0)$, $(-1,0,0)$, $(-2,0,0)$, $(0,0,-2)$ \\
\cline{2-3}
& $\pm n^{\prime } \cdot (0,0,1)$
& $(0,0,0)$, $(-1,0,0)$, $(-2,0,0)$, $(0,0,-2)$ \\
\cline{2-3}
& $\pm n^{\prime } \cdot (1,0,1)$
& $(0,0,0)$, $(-1,0,0)$, $(-2,0,0)$, $(0,0,-2)$ \\
\cline{2-3}
& $\pm n^{\prime } \cdot (1,0,-1)$
& $(0,0,0)$, $(-1,0,0)$, $(-2,0,0)$, $(0,0,-2)$ \\
\cline{1-3}
$n\cdot (1,0,0), n=-9,\ldots,9$ & 
$\pm n^{\prime } \cdot (0,1,0)$
& $(0,-2,0)$, $(0,0,-2)$ \\
\cline{2-3}
& $\pm n^{\prime } \cdot (0,0,1)$
& $(0,-2,0)$, $(0,0,-2)$ \\
\cline{2-3}
& $\pm n^{\prime } \cdot (0,1,1)$
& $(0,-2,0)$, $(0,0,-2)$ \\
\cline{2-3}
& $\pm n^{\prime } \cdot (0,1,-1)$
& $(0,-2,0)$, $(0,0,-2)$ \\
\cline{1-3}
$n\cdot (0,1,1), n=-4,\ldots,4$ & $\pm n^{\prime } \cdot (1,0,0)$ &
$(0,0,0)$, $(-1,0,0)$, $(-2,0,0)$ \\
\cline{1-3}
$n\cdot (0,1,-1), n=-4,\ldots,4$ & $\pm n^{\prime } \cdot (1,0,0)$ &
$(0,0,0)$, $(-1,0,0)$, $(-2,0,0)$, $(-1,-1,-1)$ \\
\cline{2-3}
& $\pm n^{\prime } \cdot (0,1,1)$ &
$(-1,-1,-1)$ \\
\cline{1-3}
$n\cdot (1,0,1), n=-4,\ldots,4$ & $\pm n^{\prime } \cdot (0,1,0)$ &
$(0,-2,0)$ \\
\cline{1-3}
$n\cdot (1,0,-1), n=-4,\ldots,4$ & $\pm n^{\prime } \cdot (0,1,0)$ &
$(0,-2,0)$, $(-1,-1,-1)$ \\
\cline{2-3}
& $\pm n^{\prime } \cdot (1,0,1)$ &
$(-1,-1,-1)$ \\
\cline{1-3}
$n\cdot (1,1,0), n=-4,\ldots,4$ & $\pm n^{\prime } \cdot (0,0,1)$ &
$(0,0,-2)$ \\
\cline{1-3}
$n\cdot (1,-1,0), n=-4,\ldots,4$ & $\pm n^{\prime } \cdot (0,0,1)$ &
$(0,0,-2)$, $(-1,-1,0)$, $(-1,-1,-1)$ \\
\cline{2-3}
& $\pm n^{\prime } \cdot (1,1,0)$ &
$(-1,-1,0)$, $(-1,-1,-1)$ \\
\cline{1-3}
$n\cdot (0,2,1), n=-3,\ldots,3$ & $\pm n^{\prime } \cdot (1,0,0)$ &
$(0,0,0)$, $(-1,0,0)$, $(-2,0,0)$ \\
\cline{1-3}
$n\cdot (0,2,-1), n=-3,\ldots,3$ & $\pm n^{\prime } \cdot (1,0,0)$ &
$(0,0,0)$, $(-1,0,0)$, $(-2,0,0)$ \\
\cline{1-3}
$n\cdot (0,1,2), n=-3,\ldots,3$ & $\pm n^{\prime } \cdot (1,0,0)$ &
$(0,0,0)$, $(-1,0,0)$, $(-2,0,0)$ \\
\cline{1-3}
$n\cdot (0,1,-2), n=-3,\ldots,3$ & $\pm n^{\prime } \cdot (1,0,0)$ &
$(0,0,0)$, $(-1,0,0)$, $(-2,0,0)$ \\
\cline{1-3}
$n\cdot (2,0,1), n=-3,\ldots,3$ & $\pm n^{\prime } \cdot (0,1,0)$ &
$(0,-2,0)$ \\
\cline{1-3}
$n\cdot (2,0,-1), n=-3,\ldots,3$ & $\pm n^{\prime } \cdot (0,1,0)$ &
$(0,-2,0)$ \\
\cline{1-3}
$n\cdot (1,0,2), n=-3,\ldots,3$ & $\pm n^{\prime } \cdot (0,1,0)$ &
$(0,-2,0)$ \\
\cline{1-3}
$n\cdot (1,0,-2), n=-3,\ldots,3$ & $\pm n^{\prime } \cdot (0,1,0)$ &
$(0,-2,0)$ \\
\cline{1-3}
$n\cdot (2,1,0), n=-3,\ldots,3$ & $\pm n^{\prime } \cdot (0,0,1)$ &
$(0,0,-2)$ \\
\cline{1-3}
$n\cdot (2,-1,0), n=-3,\ldots,3$ & $\pm n^{\prime } \cdot (0,0,1)$ &
$(0,0,-2)$ \\
\cline{1-3}
$n\cdot (1,2,0), n=-3,\ldots,3$ & $\pm n^{\prime } \cdot (0,0,1)$ &
$(0,0,-2)$ \\
\cline{1-3}
$n\cdot (1,-2,0), n=-3,\ldots,3$ & $\pm n^{\prime } \cdot (0,0,1)$ &
$(0,0,-2)$, $(-2,-1,0)$ \\
\cline{2-3}
& $\pm n^{\prime } \cdot (2,1,0)$ &
$(-2,-1,0)$ \\
\hline
\end{tabular}
\caption{Sets of staple-shaped gauge link paths and spatial pion momenta
${\bf P} $ used on the lattice. Gauge link paths are characterized by
the quark separation vector ${\bf b} $ and the staple vector
$\eta {\bf v} $, cf.~Fig.~\ref{fig_2a}. The surveyed range of $\eta $,
parameterized in the table by the integer $n^{\prime } $, was always
chosen to extend from zero to well beyond the point where a numerical
signal ceases to be discernible. The maximal magnitude of the
Collins-Soper parameter $\hat{\zeta} $ attained in these sets is
$|\hat{\zeta} |=2.03$, for ${\bf P} \cdot aL/(2\pi)=(-2,0,0)$ paired
with $\eta {\bf v} /a =\pm n^{\prime } \cdot (1,0,0)$ and rotations
thereof.}
\label{stapletable}
\end{table}

\section{Numerical results}
\subsection{SIDIS and DY limits}
\begin{figure}
\psfig{file=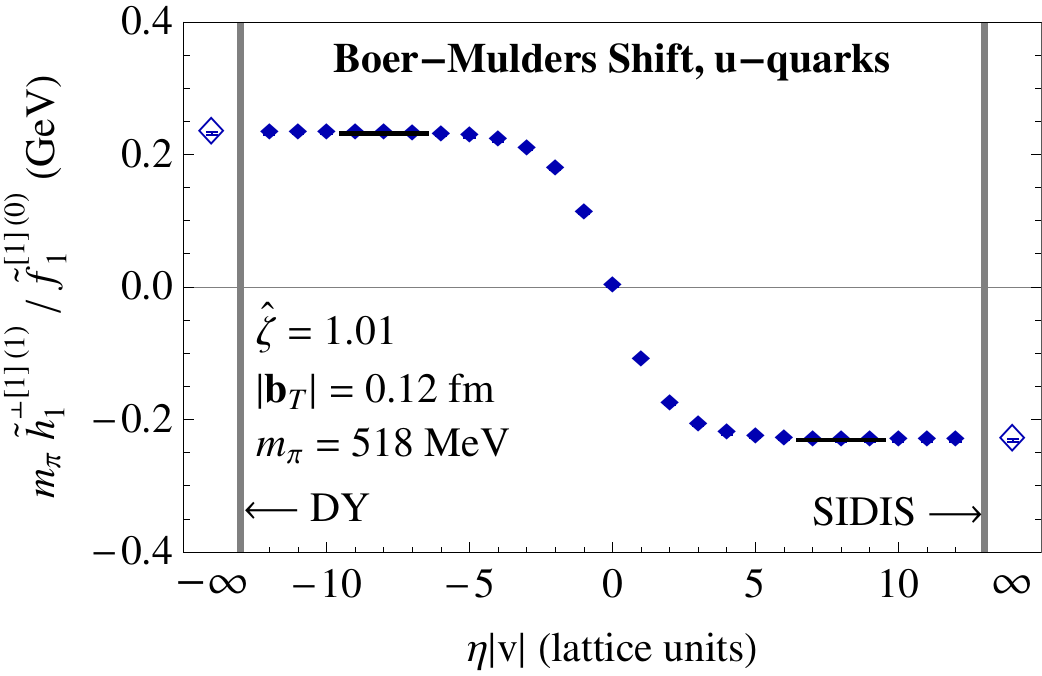,width=8.5cm}
\hfill
\psfig{file=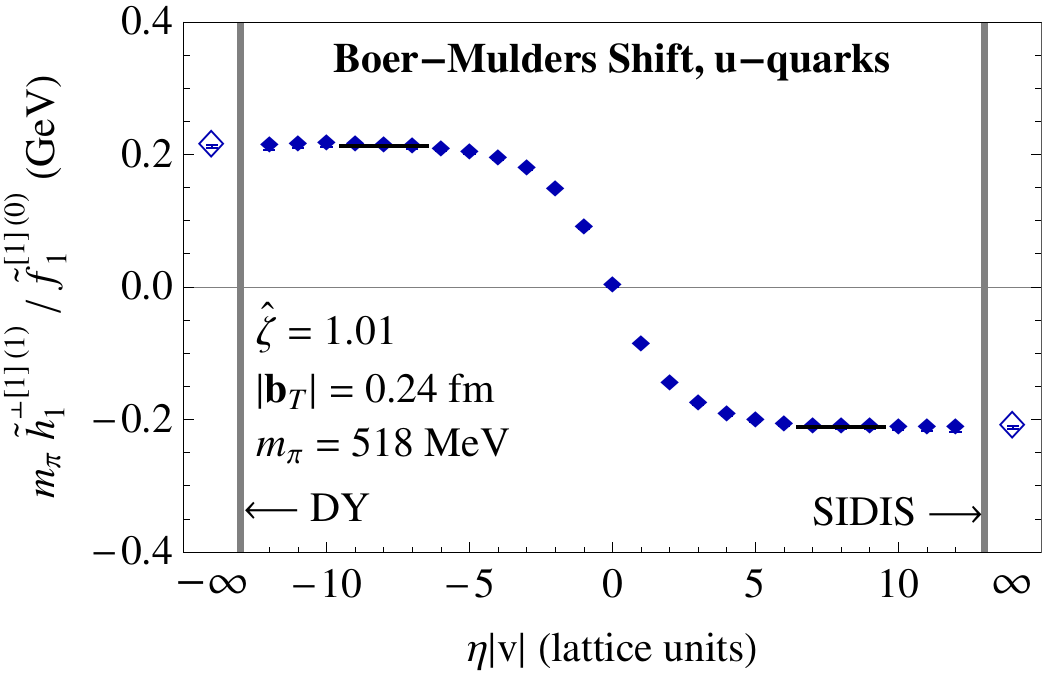,width=8.5cm}
\\
\psfig{file=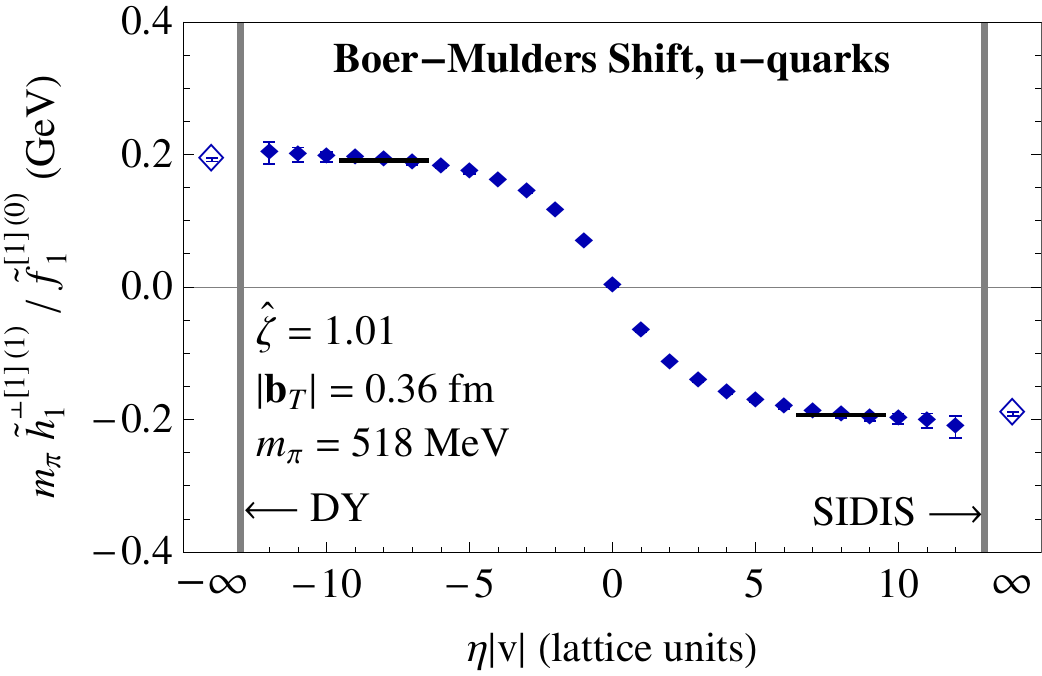,width=8.5cm}
\hfill
\psfig{file=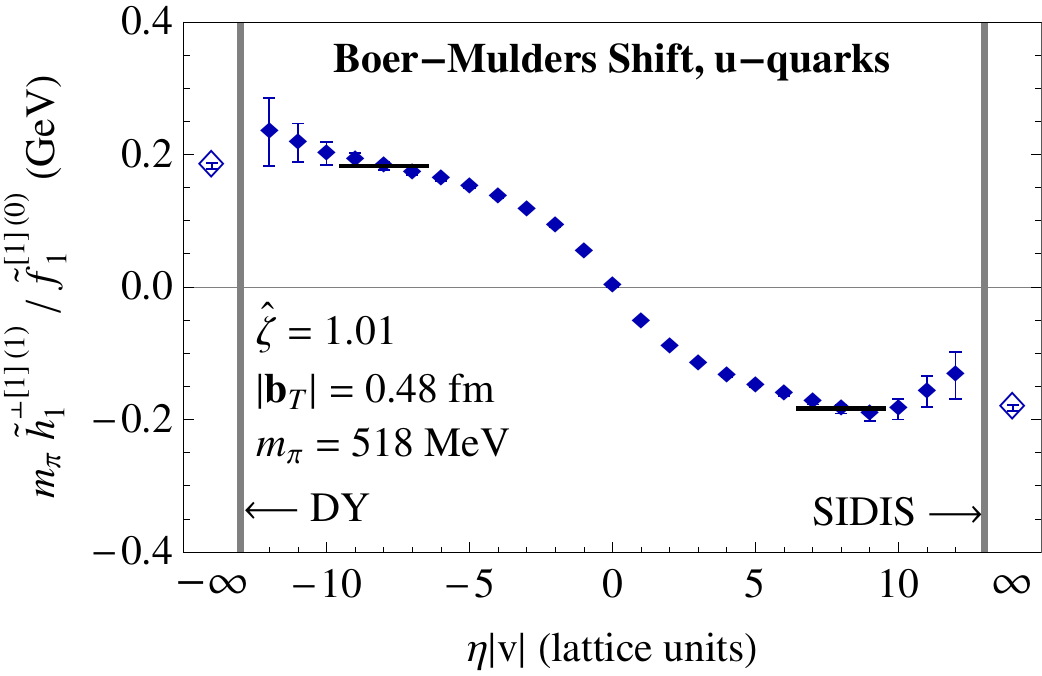,width=8.5cm}
\caption{Generalized Boer-Mulders shift for $u$-quarks as a function of
staple extent $\eta |v|$, for fixed $\hat{\zeta } =1.01$; the panels
illustrate data obtained at a succession of quark separations $|b_T |$.
Plateau fits and extraction of asymptotic values (open symbols) are
described in the main text.}
\label{vseta1}
\end{figure}

The first step in the analysis of the obtained data concerns the behavior
as a function of staple length $\eta $. For ease of notation, both positive
and negative $\eta $ are considered for a fixed $v\cdot P > 0$ to
distinguish staples oriented in the forward and backward directions with
respect to the pion momentum. Of particular physical interest is the
asymptotic behavior for $\eta \rightarrow \pm \infty $, corresponding to
the SIDIS and DY limits. Fig.~\ref{vseta1} displays results for the
$u$-quark generalized Boer-Mulders shift as a function of $\eta |v|$ at a
fixed $\hat{\zeta } =1.01$, with each of the four panels corresponding
to a successively larger transverse quark separation $|b_T |$. The T-odd
behavior of the observable is evident. As the SIDIS and DY limits are
approached, a clear plateau behavior in $\eta |v|$ is observed up to
moderate values of $|b_T |$; as $|b_T |$ rises, statistical uncertainties
increase (as indicated by the jackknife error estimates in the plots),
and the identification of the plateaus becomes more tenuous, cf., e.g.,
the lower right panel, for $|b_T |=0.48\, \mbox{fm} $. Plateau values
are extracted by averaging over the regions $7a \leq \eta |v| \leq 9a$
and $-7a \geq \eta |v| \geq -9a$, respectively, as indicated by the fit
lines in the plots; finally, the SIDIS and DY limits are obtained imposing
T-oddness, i.e., the two plateau values in each plot are averaged with a
relative minus sign to yield the asymptotic SIDIS and DY estimates also
displayed in the panels (open symbols). The asymptotic values slightly
decrease in magnitude as $|b_T |$ rises.

\begin{figure}
\centerline{\psfig{file=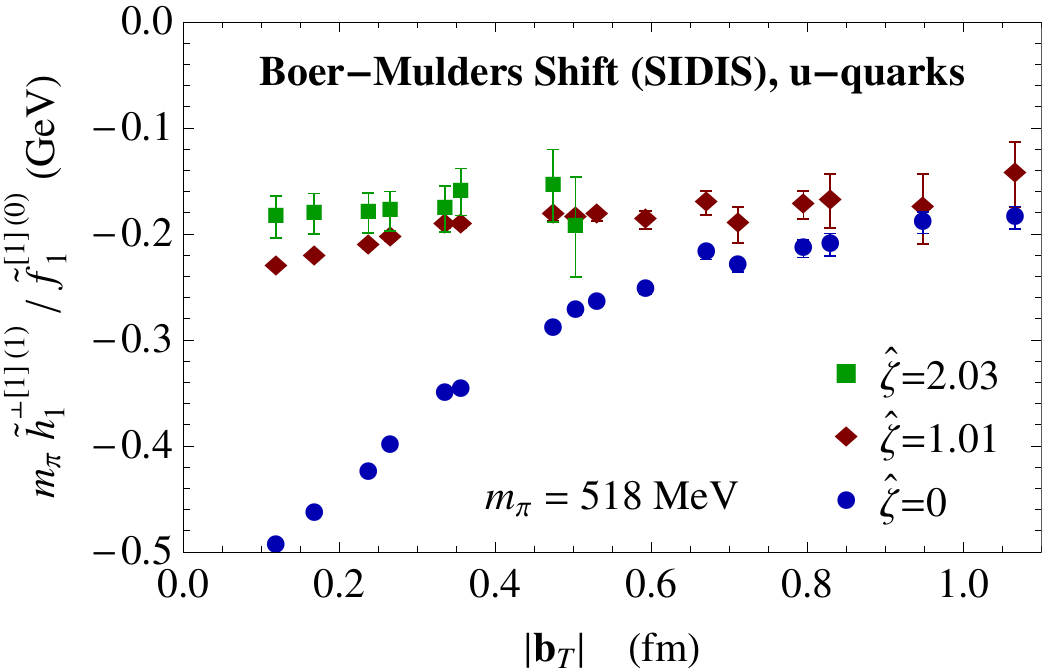,width=8.5cm} }
\caption{Generalized Boer-Mulders shift for $u$-quarks in the SIDIS limit
as a function of the quark separation $|b_T |$, for three different
values of the Collins-Soper parameter $\hat{\zeta } $. The shown
uncertainties are statistical jackknife errors.}
\label{vsb}
\end{figure}

Fig.~\ref{vsb} summarizes the results obtained in the SIDIS limit
as a function of the quark separation $|b_T |$, for three different
values of the Collins-Soper parameter $\hat{\zeta } $. Note that
the data at small $|b_T |$, up to $|b_T | \approx 2a = 0.24\, \mbox{fm} $,
may be affected by discretization artefacts, but at larger $|b_T |$,
the data are expected to well approximate the continuum limit. For
larger $\hat{\zeta } $, cf.~also further examples below, the
statistical fluctuations rapidly increase, and no useful signal
was obtained beyond $|b_T | =0.5\, \mbox{fm} $ in the case
$\hat{\zeta } =2.03$. The data appear to approach well-defined
limits as either $|b_T |$ or $\hat{\zeta } $ becomes large. The
behavior as a function of $\hat{\zeta } $ will be discussed in
greater detail below; the behavior as $|b_T |$ becomes large seems
plausible: Physically, once $|b_T |$ exceeds the size of the pion,
the correlator (\ref{spacecorr}) cannot anymore probe correlations
inside the pion; it only contains vacuum-vacuum and vacuum-pion
correlations. The $|b_T |$-dependence of these correlations is then
expected to be dominated by the typical exponential fall-off with
$|b_T |$ observed in the vacuum\footnote{Heuristically, the expectation
value of the gauge link staple, which, after integrating out the quark
fields, may be thought of as being completed into a closed loop by the
(fluctuating) world lines of dynamically propagating quarks, is expected
to be determined by the chromodynamic flux piercing the loop. This is,
e.g., the origin of the Wilson loop area law demonstrating confinement
in Yang-Mills theory \cite{greensite}. Once $|b_T |$ exceeds the size
of the pion, at most one of the legs of the staple can traverse the pion;
the other runs entirely within the vacuum. Consider now varying $|b_T |$
by shifting the latter leg; the area being added or removed from the
loop lies purely within the vacuum. Only vacuum chromodynamic flux is
being added or subtracted, while the chromodynamic flux influenced by
the pion remains fixed. Thus, the variation of the expectation value
with $|b_T |$ is determined purely by vacuum properties (which is not
to say that the expectation value becomes entirely independent of the
properties of the pion; only its $|b_T |$-dependence does). This
argument is unchanged if one subsequently averages over different
positions of the leg of the staple traversing the pion.}; neglecting
all other dependences in comparison, and canceling the dominant
behavior in the ratio (\ref{gbmshift}) leads to the expectation
of a constant asymptotic behavior in $|b_T |$.

As $\hat{\zeta } $ increases, one furthermore would expect the
aforementioned constant to converge to an asymptotic limit. Remarkably,
however, even the $\hat{\zeta } =0$ data appear to already approach
the same large-$|b_T |$ constant as the data at higher $\hat{\zeta } $.
Possibly, this may be understood as a consequence of the Lorentz
invariance of the vacuum; only vacuum-vacuum and vacuum-pion
correlations are probed at large $|b_T |$, and it seems plausible
that these would be independent of the pion momentum. It would be
desirable to gain a more definite understanding of this property.
The most remarkable feature of the data, however, is the apparent
tendency of the generalized Boer-Mulders shift to become constant
in $|b_T |$ as $\hat{\zeta } $ is increased, not only for asymptotic
values of $|b_T |$, but for all $|b_T |$. No obvious reason for this
behavior at low to intermediate $|b_T |$ is apparent, and it would be
very interesting to develop an understanding of it. The constant
behavior implies that, in the relevant transverse momentum $|k_T |$ range
corresponding to the probed range of $|b_T |$, the transverse momentum
spectrum of polarized quarks is the same as for unpolarized ones.

\subsection{Evolution in $\hat{\zeta } $}
\label{zetasec}
\begin{figure}
\psfig{file=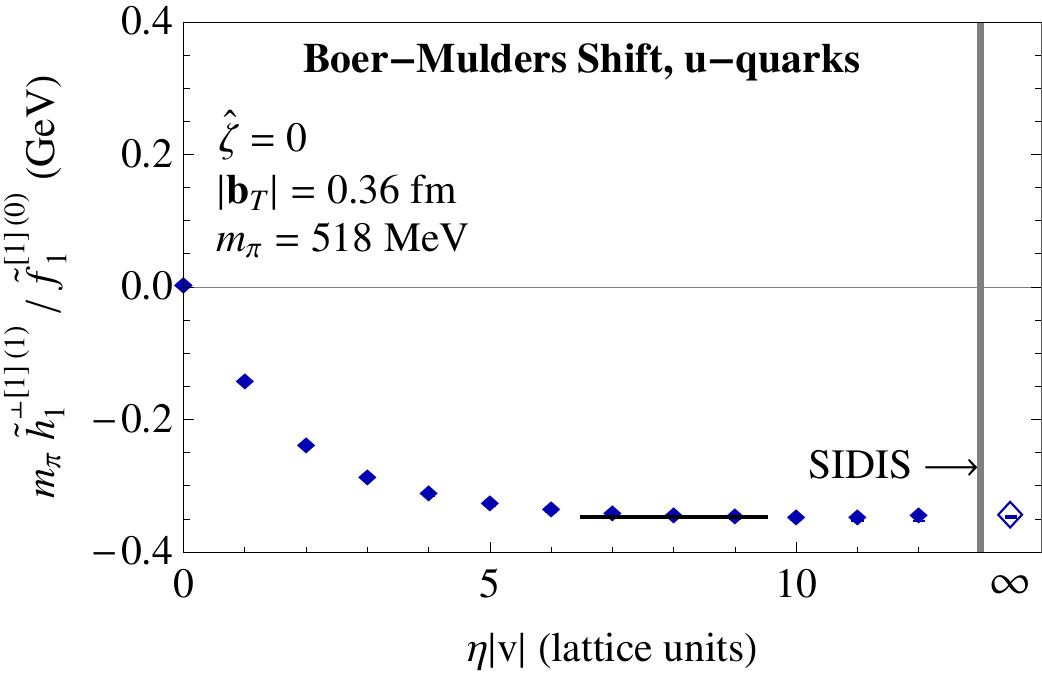,width=8.5cm}
\hfill
\psfig{file=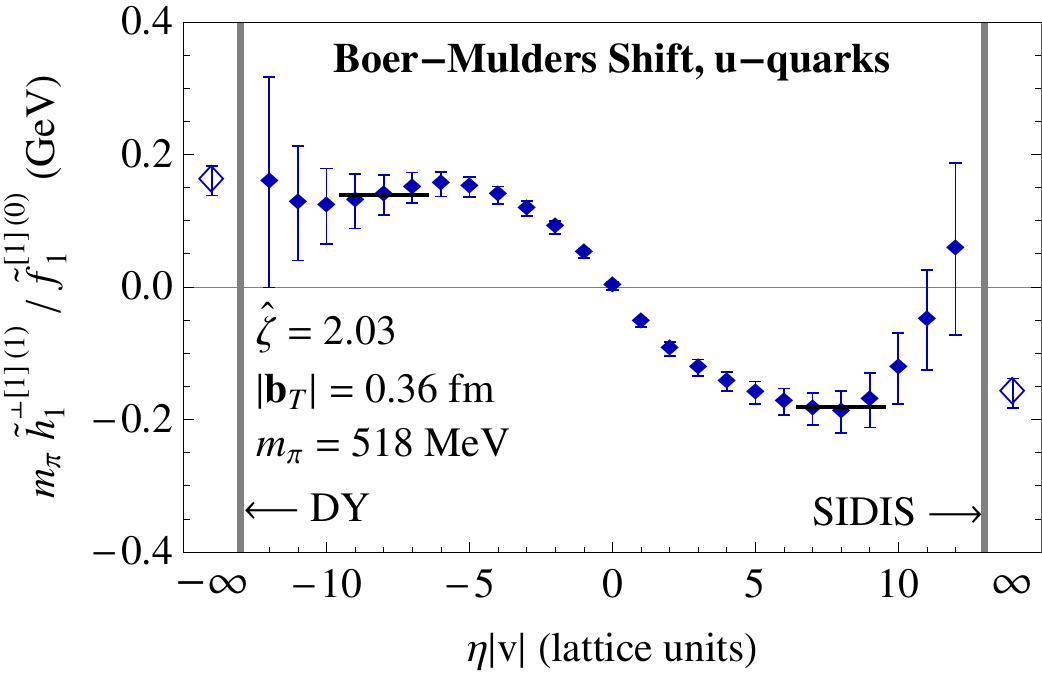,width=8.5cm}
\caption{Generalized Boer-Mulders shift for $u$-quarks as a function of
staple extent $\eta |v|$, analogous to Fig.~\ref{vseta1}, but at fixed
$|b_T | =0.36\, \mbox{fm} $ and $\hat{\zeta } $ varying between the panels
instead. In terms of a $\hat{\zeta } $ sequence, the lower left panel of
Fig.~\ref{vseta1} lies in between the two panels shown here. For pion
spatial momentum ${\bf P} =0$ (left panel), there is only one branch as
a function of $\eta |v|$.}
\label{vseta2}
\end{figure}

A special focus of the present investigation is the behavior as a function
of $\hat{\zeta } $ and the large $\hat{\zeta } $ limit, i.e., studying
in detail the behavior of the sequence of data seen in Fig.~\ref{vsb}
at a fixed value of $|b_T |$. Fig.~\ref{vseta2} shows two further panels
analogous to the ones in Fig.~\ref{vseta1}, but with $\hat{\zeta } $
varying between the panels and $|b_T | =0.36\, \mbox{fm} $ fixed instead.
Thus, in terms of a $\hat{\zeta } $ sequence, the lower left panel of
Fig.~\ref{vseta1} lies in between the two panels displayed in
Fig.~\ref{vseta2}. As already mentioned further above, for pion spatial
momentum ${\bf P} =0$, corresponding to the left panel in Fig.~\ref{vseta2},
there is only one branch as a function of $\eta |v|$, as shown. The right
panel in Fig.~\ref{vseta2}, corresponding to $\hat{\zeta } =2.03$,
illustrates the rapid deterioration of signal as the pion momentum is
increased. Nevertheless, at the moderate $|b_T |$ used here, a plateau
can still be extracted.

\begin{figure}
\psfig{file=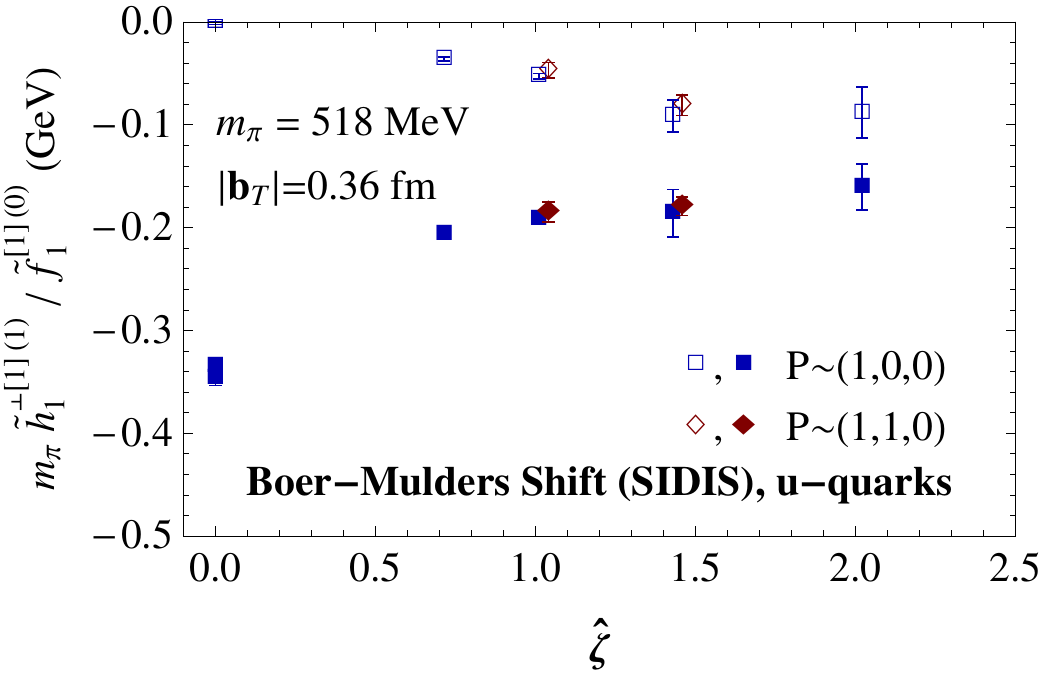,width=8.5cm}
\hfill
\psfig{file=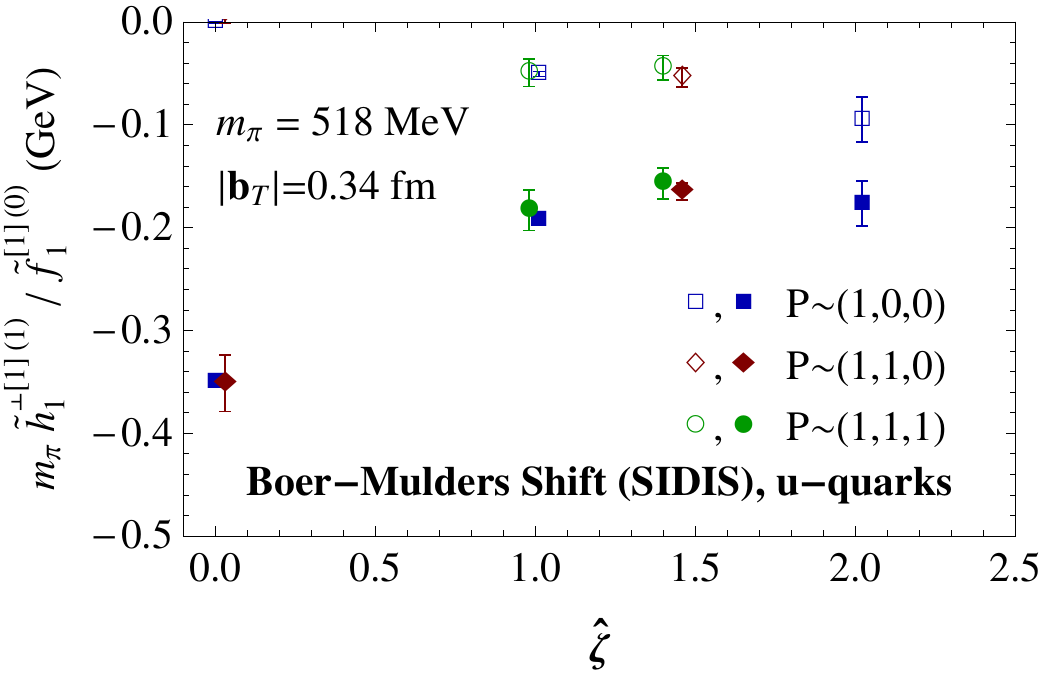,width=8.5cm}
\caption{Generalized Boer-Mulders shift for $u$-quarks in the SIDIS limit
as a function of the Collins-Soper parameter $\hat{\zeta } $, for fixed
quark separation $|b_T | =0.36\, \mbox{fm} $ (left) and
$|b_T | =0.34\, \mbox{fm} $ (right). Filled symbols correspond to
the full generalized Boer-Mulders shift, whereas open symbols correspond
to the partial contribution obtained by replacing $\widetilde{A}_{4B} $
in (\ref{gbmshift}) by $\widetilde{A}_{4} $, for further discussion,
cf.~main text. Spatial pion momenta ${\bf P} $ along various directions,
as labeled, were employed; labels in general subsume more than one
magnitude along the specified axis, cf.~Table~\ref{stapletable} for the
complete set of momenta for which data were generated. Data labeled
${\bf P} \sim (1,0,0)$ include averaging over 90 degree rotations
thereof, i.e., momenta along any of the three spatial lattice axes
are included. A given value of $\hat{\zeta } $ can be accessed using
different pairs of ${\bf P} $ and staple direction $v$; in such cases,
data corresponding to different ${\bf P} $ directions are slightly
offset from one another horizontally for better visibility.
The shown uncertainties are statistical jackknife errors.}
\label{vszeta3x}
\end{figure}

Fig.~\ref{vszeta3x} summarizes all results obtained in the SIDIS limit
at $|b_T | =0.36\, \mbox{fm} $ as a function of the Collins-Soper
parameter $\hat{\zeta } $ (left panel), and analogous data for the
nearby value $|b_T | =0.34\, \mbox{fm} $ (right panel).
From the figure, the good rotational properties of the calculation
are evident; a given value of $\hat{\zeta } $ can
be accessed using different directions of the pion spatial momentum
${\bf P} $ and the staple direction $v$, including both on- and off-axis
directions, as shown. The corresponding results coincide, indicating
that potential lattice artefacts are under control, and thus buttressing
the physical significance of the data obtained. To assess the asymptotic
behavior at large $\hat{\zeta } $, it is advantageous to consider not
only the full generalized Boer-Mulders shift, but also the partial
contribution obtained by replacing $\widetilde{A}_{4B} $ in
(\ref{gbmshift}) by $\widetilde{A}_{4} $, omitting the contribution
from $\widetilde{B}_{3} $, cf.~eq.~(\ref{a4bdef}). Both quantities
are displayed Fig.~\ref{vszeta3x}; as is evident from the figure,
the partial $\widetilde{A}_{4} $ contribution vanishes at
$\hat{\zeta } =0$, but monotonically increases in magnitude as
$\hat{\zeta } $ is increased. By contrast, the remaining contribution
from $\widetilde{B}_{3} $ to the full generalized Boer-Mulders shift,
dominant at $\hat{\zeta } =0$, decreases in magnitude as $\hat{\zeta } $ 
rises. This matches the behavior expected from eq.~(\ref{a4bdef}),
according to which the contribution from $\widetilde{B}_{3} $ becomes
insignificant for $\hat{\zeta } \rightarrow \infty $ under the assumption
that the amplitude $\widetilde{B}_{3} $ stays finite. Altogether, the
full generalized Boer-Mulders shift also decreases in magnitude as
$\hat{\zeta } $ increases. Thus, the behavior of the data suggests that,
by considering both quantities, one has access to both lower and upper
bounds for the generalized Boer-Mulders shift, considerably increasing
the confidence in the extrapolations to large $\hat{\zeta } $ discussed
below. The comparison between the partial and full quantities also permits
an assessment of the extent to which evolution in $\hat{\zeta } $ has
progressed towards the asymptotic limit. Evidently, according to
Fig.~\ref{vszeta3x}, already about half of the magnitude of the full
generalized Boer-Mulders shift is subsumed in the partial
$\widetilde{A}_{4} $ contribution at $\hat{\zeta } \approx 2$.
A significant part of the $\hat{\zeta } $ evolution has thus already
been achieved at that value of $\hat{\zeta } $. In this respect, the
present study yields a much clearer picture than was obtained in the
previous nucleon investigation \cite{tmdlat}.

\begin{figure}
\psfig{file=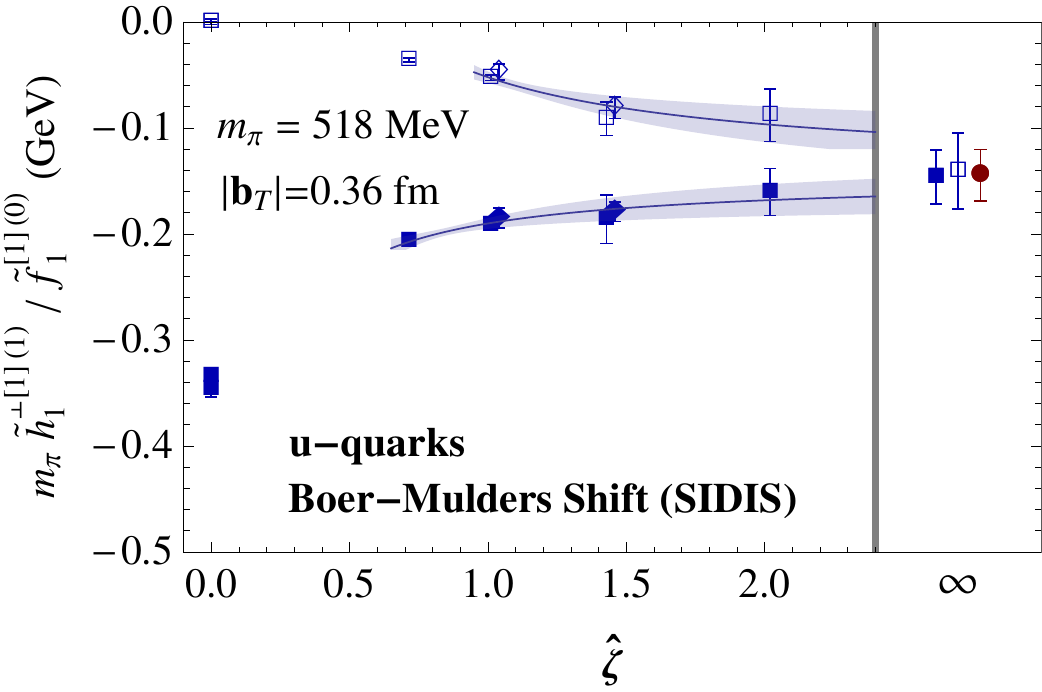,width=8.5cm}
\hfill
\psfig{file=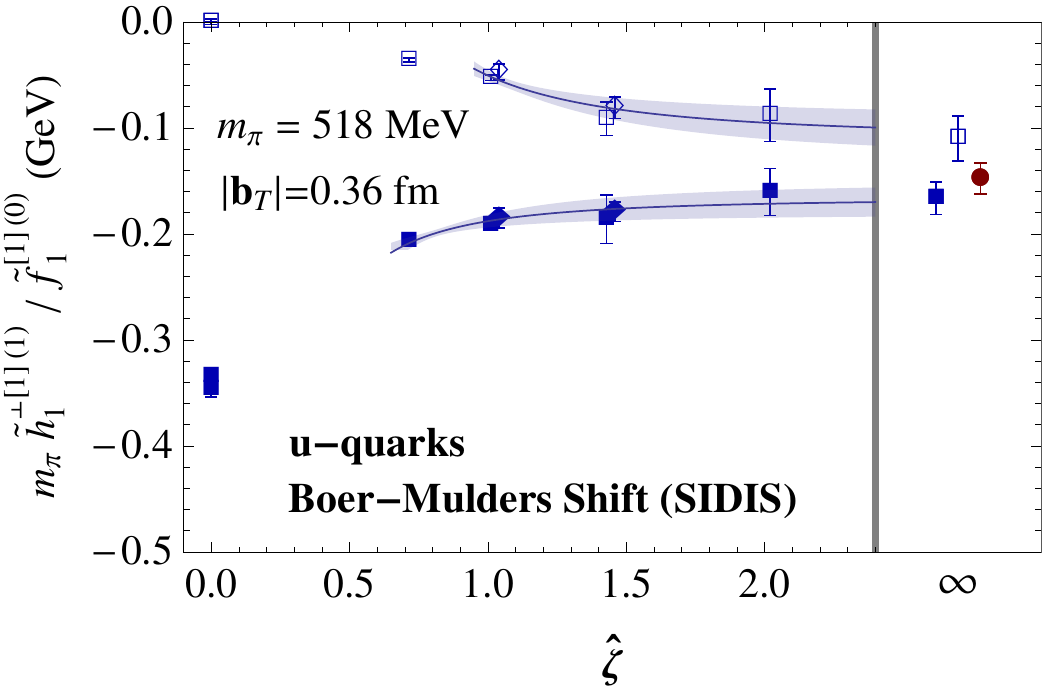,width=8.5cm}
\caption{Least-squares fits to the generalized Boer-Mulders shift data
shown in Fig.~\ref{vszeta3x} (left). Separate fits to filled and open
data points are shown. Left panel displays fits obtained
with ansatz $c+d/\hat{\zeta } $; right panel displays fits obtained with
ansatz $c+d/\hat{\zeta }^{2} $. Error bands show statistical jackknife
uncertainties of the fits. Data points that lie outside the horizontal
range of the displayed fit curves were not included in fits; in the
case of the open symbols, one less data point was taken into account
because it already lies in the region of an apparent inflection point,
at which the simple asymptotic fit ansatz certainly is not applicable
anymore. The asymptotic values displayed include the separate fits to
the filled and open data points (filled and open squares), as well as
the results of combined fits (filled circles); for details, cf.~main text.
The shown uncertainties are statistical jackknife errors.}
\label{fit1}
\end{figure}

\begin{figure}
\psfig{file=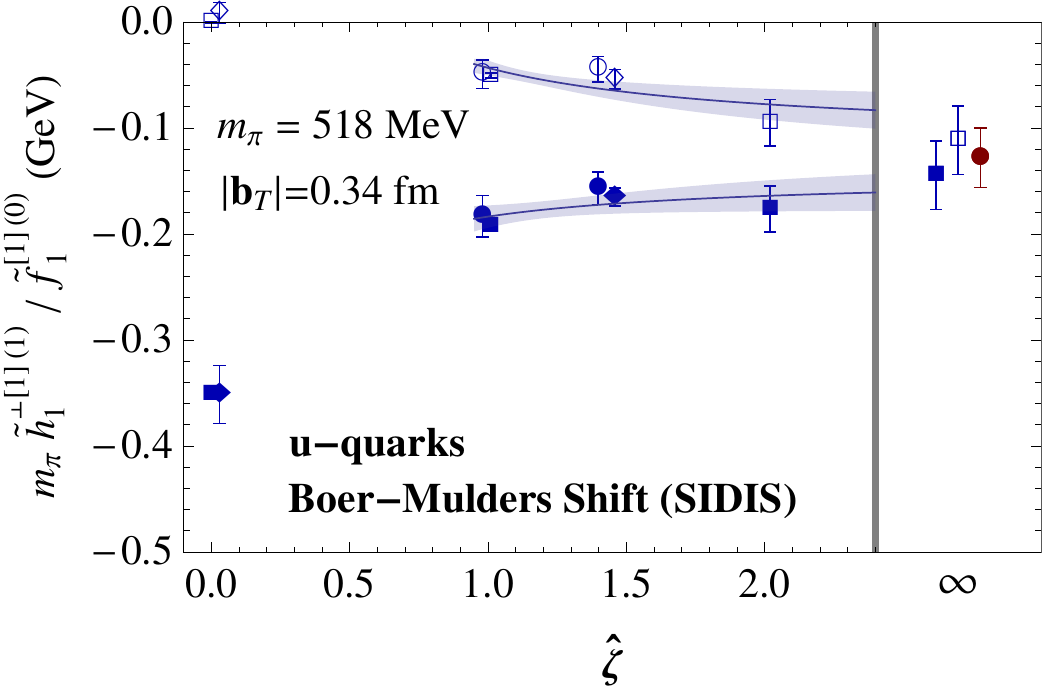,width=8.5cm}
\hfill
\psfig{file=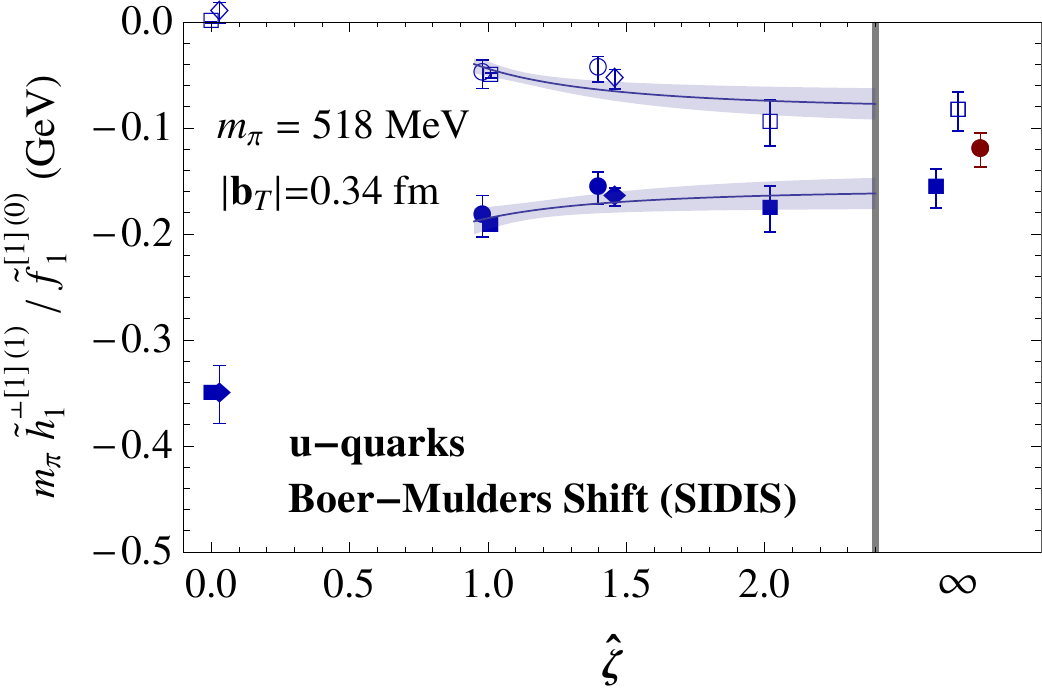,width=8.5cm}
\caption{Least-squares fits to the generalized Boer-Mulders shift data
shown in Fig.~\ref{vszeta3x} (right). Separate fits to filled and open
data points are shown. Left panel displays fits obtained
with ansatz $c+d/\hat{\zeta } $; right panel displays fits obtained with
ansatz $c+d/\hat{\zeta }^{2} $. Error bands show statistical jackknife
uncertainties of the fits. Data points that lie outside the horizontal
range of the displayed fit curves were not included in fits. The
asymptotic values displayed include the separate fits to the filled
and open data points (filled and open squares), as well as the results
of combined fits (filled circles); for details, cf.~main text. The shown
uncertainties are statistical jackknife errors.}
\label{fit2}
\end{figure}

\begin{figure}
\psfig{file=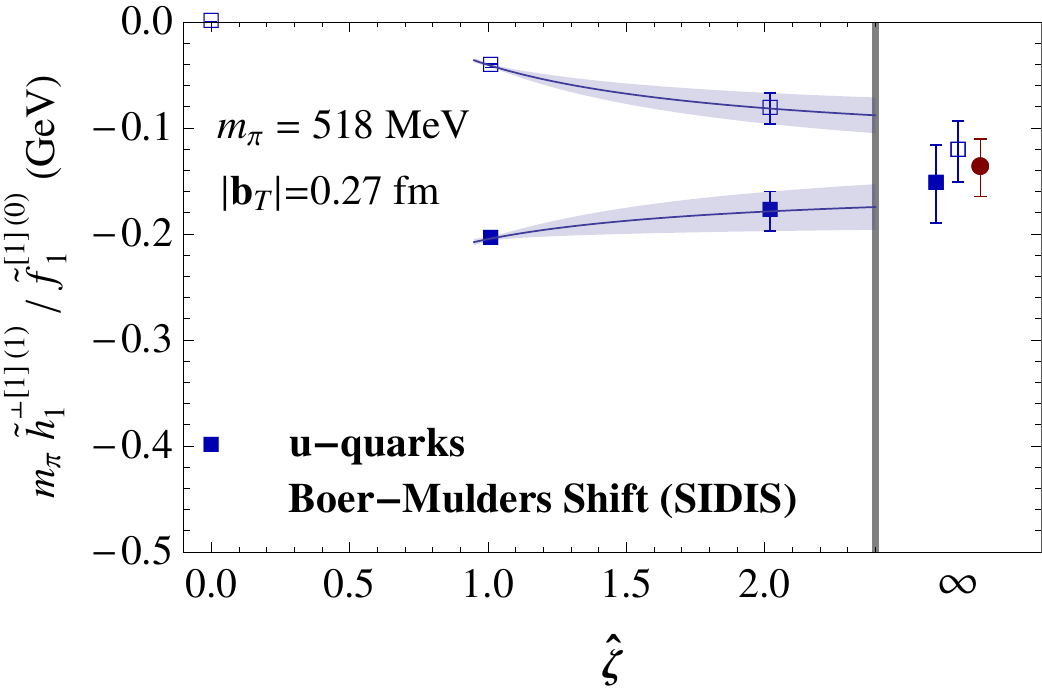,width=8.5cm}
\hfill
\psfig{file=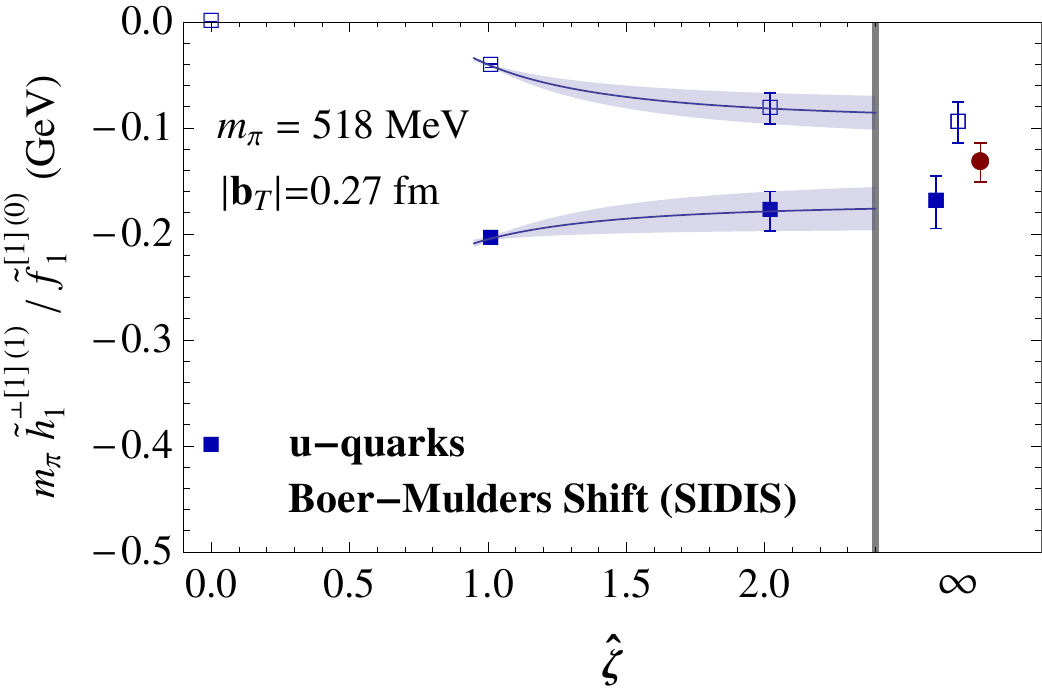,width=8.5cm}
\caption{Least-squares fits to generalized Boer-Mulders shift data at
$|b_T | =0.27\, \mbox{fm} $, analogous to Figs.~\ref{fit1} and \ref{fit2}.
The data were all obtained using spatial pion momenta ${\bf P} $ along
the lattice axes (${\bf P} \sim (1,0,0)$ in the nomenclature of
Fig.~\ref{vszeta3x}, including multiples and 90 degree rotations thereof,
cf.~Table~\ref{stapletable} for the complete set of momenta for which
data were generated). Separate fits to filled and open
data points are shown. Left panel displays fits obtained with ansatz
$c+d/\hat{\zeta } $; right panel displays fits obtained with ansatz
$c+d/\hat{\zeta }^{2} $. Error bands show statistical jackknife
uncertainties of the fits. Data points that lie outside the horizontal
range of the displayed fit curves were not included in fits. The
asymptotic values displayed include the separate fits to the filled
and open data points (filled and open squares), as well as the results
of combined fits (filled circles); for details, cf.~main text. The
shown uncertainties are statistical jackknife errors.}
\label{fit3}
\end{figure}

To obtain quantitative statements about the large $\hat{\zeta } $ limit,
least-squares fits\footnote{Whereas the data at higher $\hat{\zeta } $
generally display the larger statistical uncertainties, the data at
lower $\hat{\zeta } $ are likely to deviate systematically to a larger
degree from any putative simple asymptotic behavior. It therefore
does not seem appropriate to bias the fits towards the lower
$\hat{\zeta } $ data by performing a $\chi^{2} $ fit taking into
account the statistical uncertainties.} to the $\hat{\zeta } $-dependences
of the data were performed, using power-correction ans\"atze of the form
$c+d/\hat{\zeta } $ and $c+d/\hat{\zeta }^{2} $. It should be noted that
the values of $\hat{\zeta } $ for which data were obtained in the present
study do not reach clearly into the perturbative regime within which a
reliable connection to perturbative evolution can be guaranteed. The
aforementioned fit functions should therefore be regarded as ad hoc
ans\"atze. On the one hand, separate fits to the full generalized
Boer-Mulders shift data and to the partial $\widetilde{A}_{4} $
contribution were performed; on the other hand, a combined fit to
both quantities with a common constant $c$ (but, of course, separate
coefficients $d$) was carried out. The results are shown in
Figs.~\ref{fit1}-\ref{fit3}, and the asymptotic values summarized
in Table~\ref{fittab}. Fig.~\ref{fit1} displays the fits to the data
of Fig.~\ref{vszeta3x} (left), at $|b_T | =0.36\, \mbox{fm} $, whereas
Fig.~\ref{fit2} displays the fits to the data of Fig.~\ref{vszeta3x}
(right), at $|b_T | =0.34\, \mbox{fm} $. Fig.~\ref{fit3} additionally
displays fits to data obtained at a lower $|b_T |$, namely,
$|b_T | =0.27\, \mbox{fm} $.

\begin{table}
\begin{tabular}{|c|c||c|c|c|c|}
\hline
& Fit function
& Full B.-M. shift & Contribution $\widetilde{A}_{4} $ &
Combined fit & RMS deviation of \\
& & (GeV) & only (GeV) & (GeV) & combined fit (GeV) \\
\hline\hline
$|b_T |=0.36\, \mbox{fm} $ & $c+d/\hat{\zeta } $ &
-0.146(26) & -0.141(36) & -0.145(25) & 0.00755 \\
\hline
$|b_T |=0.36\, \mbox{fm} $ & $c+d/\hat{\zeta }^{2} $ &
-0.166(16) & -0.110(22) & -0.148(15) & 0.01695 \\
\hline\hline
$|b_T |=0.34\, \mbox{fm} $ & $c+d/\hat{\zeta } $ &
-0.145(33) & -0.112(33) & -0.128(29) & 0.01466 \\
\hline
$|b_T |=0.34\, \mbox{fm} $ & $c+d/\hat{\zeta }^{2} $ &
-0.157(19) & -0.084(19) & -0.121(16) & 0.02315 \\
\hline\hline
$|b_T |=0.27\, \mbox{fm} $ & $c+d/\hat{\zeta } $ &
-0.153(37) & -0.122(29) & -0.138(28) & 0.00975 \\
\hline
$|b_T |=0.27\, \mbox{fm} $ & $c+d/\hat{\zeta }^{2} $ &
-0.170(25) & -0.095(20) & -0.133(19) & 0.03855 \\
\hline
\end{tabular}
\caption{Asymptotic values of least-squares fits to the
$\hat{\zeta } $-dependences of the full generalized Boer-Mulders
shift and the partial $\widetilde{A}_{4} $ contribution separately,
as well as a combined fit. These are the values displayed at
$\hat{\zeta } =\infty $ in Figs.~\ref{fit1}-\ref{fit3}. Uncertainties
quoted are statistical jackknife errors. The right-hand column gives
the root mean square deviation, per degree of freedom, of the combined
fit from the data means.}
\label{fittab}
\end{table}

The fits using the form $c+d/\hat{\zeta } $ are superior to those
using the form $c+d/\hat{\zeta }^{2} $. In the former case,
the asymptotic values obtained by analyzing the full generalized
Boer-Mulders shift data and the partial $\widetilde{A}_{4} $ contribution
separately agree within uncertainties, and also with the result of the
combined fit. In the latter case, the asymptotic values obtained with
the separate fits differ significantly. Not surprisingly, the results
of the combined fits using the two fit ans\"atze are quite similar;
by construction, they settle roughly in the middle between the two
separate quantities. Nevertheless, also in the combined fits, the
$c+d/\hat{\zeta } $ ansatz is more favorable, as evidenced by the
last column in Table~\ref{fittab}, which gives the root mean square
deviation, per degree of freedom, of the combined fit from the data
means. The deviation is considerably larger for the $c+d/\hat{\zeta }^{2} $
fits; it is of the order of, or larger than, the jackknife statistical
uncertainty of the asymptotic value. By contrast, for the
$c+d/\hat{\zeta } $ ansatz, the deviation is considerably smaller
than the jackknife statistical uncertainty of the asymptotic value.
These observations favor the $c+d/\hat{\zeta } $ fit ansatz.

Note that the asymptotic values for the generalized Boer-Mulders shift
obtained using the $c+d/\hat{\zeta } $ ansatz both at
$|b_T | =0.36\, \mbox{fm} $ and at $|b_T | =0.34\, \mbox{fm} $,
as well as at the lower separation $|b_T | =0.27\, \mbox{fm} $,
all coincide within uncertainties. These extrapolations thus do not
modify the observation made further above in connection with
Fig.~\ref{vsb}, namely, that the generalized Boer-Mulders shift
appears to become constant in $|b_T |$ as $\hat{\zeta } $ becomes
large.

Altogether, it is encouraging to observe that the data generated
in the present study are of sufficient quality to ensure that the signal
survives extrapolation to the $\hat{\zeta } \rightarrow \infty $ limit,
with an uncertainty as low as, roughly, 20\% for the selected parameter
values analyzed above. It is possible to tentatively discriminate
between different asymptotic models. Further improvements of the
analysis are to be expected by confronting lattice data with
perturbative evolution equations for the observables considered
here.

\subsection{Pion-proton comparison}
\begin{figure}
\psfig{file=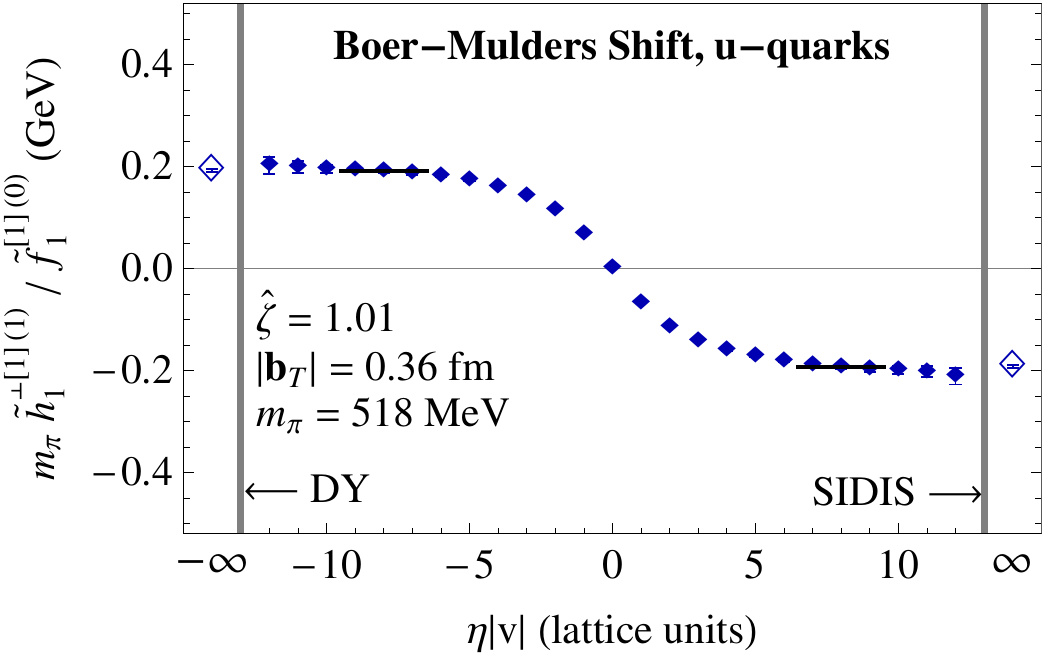,width=8.5cm}
\hspace{0.63cm}
\psfig{file=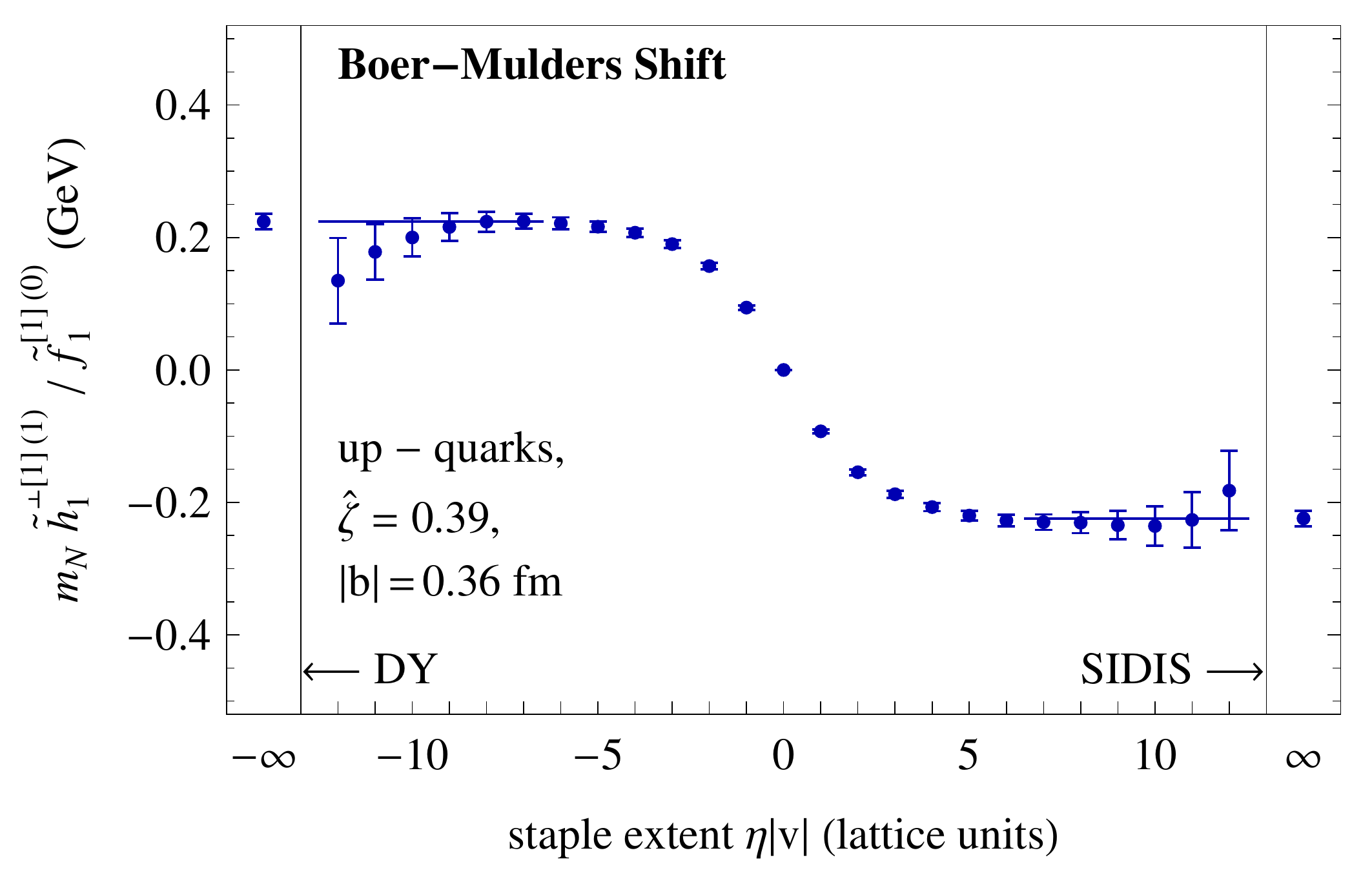,width=8.6cm}
\caption{Generalized Boer-Mulders shift for $u$-quarks in a
$\pi^{+} $-meson (left) and a proton \cite{tmdlat} (right), as a
function of staple extent $\eta |v|$, for quark separation
$|b_T | =0.36\, \mbox{fm} $. The data in the two panels were
obtained using the same spatial hadron momentum ${\bf P}$ and
staple direction $v$. This corresponds to identical
$\zeta = 2m_h \hat{\zeta } $, but, consequently, differing $\hat{\zeta } $,
as labeled in the panels, cf.~main text for further discussion.
Plateau fits and extraction of asymptotic values are analogous to
Fig.~\ref{vseta1}. In the proton case, taken from the analysis
described in \cite{tmdlat}, a wider fit range was used.}
\label{pionproton}
\end{figure}

In \cite{mb_1}, the question was posed how similar Boer-Mulders functions
are in different hadrons, and also for different flavors. The present
study permits addressing this question in passing, since Boer-Mulders
data are now available on the same lattice ensemble both for nucleons
\cite{tmdlat} and for pions, obtained here, with corresponding parameters,
facilitating a comparison. Fig.~\ref{pionproton} provides a sample
juxtaposition of the $\pi^{+} $-meson and proton $u$-quark generalized
Boer-Mulders shifts generated on the same ensemble, at identical transverse
quark separation $|b_T | =0.36\, \mbox{fm} $, and identical spatial hadron
momentum ${\bf P}$ and staple direction $v$, as a function of staple
extent. This corresponds to identical $\zeta = 2m_h \hat{\zeta } $,
but of course $\hat{\zeta } $ then differs because of the appearance
of the hadron mass in the denominator of (\ref{zetahat}). In this
particular juxtaposition, the behavior of the two shifts is seen to
be very similar. In the proton, the approach to the large $|\eta |$
plateaus appears to be slightly faster; the plateaus lie within 10\%
of one another. One could argue that the more appropriate comparison is one
at identical $\hat{\zeta } $, and not at identical $\zeta $, as shown here.
However, using identical $\hat{\zeta } $ makes the correspondence hardly
less favorable. If one interpolates the SIDIS limit pion data at
$|b_T | =0.36\, \mbox{fm} $ shown in Fig.~\ref{vszeta3x} (left) from
$\hat{\zeta } =1.01$ to $\hat{\zeta } =0.39$, the $\hat{\zeta } $ value
corresponding to the proton data in Fig.~\ref{pionproton} (right), then
the pion data are enhanced in magnitude by around 20\%, rendering the
pion data about 10\% higher in magnitude than the proton data, as
opposed to 10\% lower in the comparison at equal $\zeta $.

Of course, it should be noted that a close quantitative correspondence
to the degree observed here for the $u$-quark generalized Boer-Mulders
shifts is not generic. In the $\pi^{+} $-meson, the $\bar{d} $-quark
Boer-Mulders shift is identical to the $u$-quark one, whereas in the
proton, the $d$-quark Boer-Mulders shift is appreciably higher in
magnitude compared to the $u$-quark one \cite{tmdlat}. The corresponding
comparison would therefore not reveal a similarly close quantitative
agreement. Nevertheless, qualitative features such as the signs of the
different flavor contributions remain in correspondence.

\section{Summary and outlook}
Building on the previous study \cite{tmdlat} of transverse
momentum-dependent parton distributions in the nucleon, the present
work focused in particular on the evolution of TMD ratios such as
the generalized Boer-Mulders shift (\ref{gbmshift}) as a function
of the Collins-Soper parameter $\hat{\zeta } $, cf.~(\ref{zetahat}).
To this end, the behavior of this TMD observable in a pion was
explored; the pion, by virtue of its lower mass, allows one to
access higher $\hat{\zeta } $, and the spinless nature of the pion
facilitates spatial averaging to increase statistics. Indeed, in
contrast to the nucleon study \cite{tmdlat}, which did not
yield definite conclusions concerning the large-$\hat{\zeta } $ limit,
the present investigation provides data of sufficient quality to perform
tentative extrapolations to $\hat{\zeta } =\infty $. While the generated
data do not reach into the $\hat{\zeta } $ region in which a clear
connection to perturbative evolution equations is given, the ad hoc
extrapolations explored here lead to estimates of the
$\hat{\zeta } =\infty $ limit of the generalized Boer-Mulders shift
(\ref{gbmshift}) to within an uncertainty of as low as roughly 20\%, for
favorable values of the transverse quark separation $b_T $,
cf.~Table~\ref{fittab}. The confidence in these extrapolations is
buttressed in particular by the ability to partition the generalized
Boer-Mulders shift into two separate contributions, as discussed in
detail in section \ref{zetasec}, such that one is provided with both
an upper and a lower bound for the asymptotic behavior.

Having obtained data for the generalized Boer-Mulders shift in the pion,
the present study, together with \cite{tmdlat}, in passing also permits
a juxtaposition of pion and nucleon TMD observables. A close correspondence
between the $u$-quark generalized Boer-Mulders shifts in a $\pi^{+} $-meson
and a proton is observed, corroborating the behavior conjectured in
\cite{pion_GPD,mb_1}.

Going forward, the success in determining the characteristics of the
$\hat{\zeta } $ evolution of the generalized Boer-Mulders shift in a pion
suggests that analogous results will also be accessible for nucleons in
future lattice studies with higher statistics, focused on obtaining
signals at higher nucleon momenta. To this end, it will be useful to
explore schemes of generating nucleon interpolating fields with a
favorable overlap with higher momentum states. Further efforts currently
in progress are concerned with the universality of TMD ratios of the
type considered here under variation of the lattice discretization scheme;
this is expected to provide empirical support for the working assumption
employed here, that the lattice TMD operators are regularized and
renormalized in analogy to the corresponding continuum operators,
by multiplicative soft factors which cancel in TMD ratios. Moreover,
lattice TMD calculations are in progress at lighter pion masses,
with a view to transferring the exploratory results obtained here to the
physical light quark mass regime.

\section*{Acknowledgments}
This work benefited from fruitful discussions with T.~Bhattacharya,
D.~Boer, V.~Braun, M.~Diehl, R.~Edwards, R.~Gupta, X.~Ji, V.~Papavassiliou,
S.~Pate, A.~Prokudin, J.~Qiu, T.~Rogers, X.~Wang, B.~Yoon and J.~Zhang.
The lattice calculations performed in this work relied on the Chroma
software suite \cite{chroma} and employed computing resources provided by
the U.S.~DOE through USQCD at Jefferson Lab. Support by the
Heisenberg-Fellowship program of the DFG (P.H.), SFB/TRR-55 and the
Alexander von Humboldt-Stiftung (A.S.), as well as by the U.S.~DOE through
grants DE-FG02-96ER40965 (M.E.) and DE-FG02-94ER40818 (J.N.), and through
contract DE-AC05-06OR23177, under which Jefferson Science Associates, LLC,
operates Jefferson Laboratory (B.M.), is acknowledged.

\end{document}